\documentclass[aps, prx,twocolumn,longbibliography,superscriptaddress,showpacs,amsmath,amssymb,floatfix]{revtex4}
\usepackage[latin9]{inputenc}
\usepackage{amssymb}
\usepackage{graphicx}
\usepackage{amsmath}
\usepackage{color}
\usepackage{mathrsfs}
\usepackage{float}
\usepackage{indentfirst}
\usepackage{mathrsfs}
\usepackage{float}
\usepackage{indentfirst}
\usepackage{textcomp}
\usepackage{comment}
\usepackage{mathtools}
\usepackage{natbib,hyperref}
\usepackage{soul}

\begin{document}

\title{{Augmenting} Density Matrix Renormalization Group with Disentanglers}

\author{Xiangjian Qian}
\affiliation{Key Laboratory of Artificial Structures and Quantum Control (Ministry of Education),  School of Physics and Astronomy, Shanghai Jiao Tong University, Shanghai 200240, China}

\author{Mingpu Qin} \thanks{qinmingpu@sjtu.edu.cn}
\affiliation{Key Laboratory of Artificial Structures and Quantum Control (Ministry of Education),  School of Physics and Astronomy, Shanghai Jiao Tong University, Shanghai 200240, China}

\date{\today}

\begin{abstract}
Density Matrix Renormalization Group (DMRG) and its extensions in the form of Matrix Product States (MPS) are arguably the choice for the study of one dimensional quantum systems in the last three decades.  However, due to the limited 
entanglement encoded in the wave-function ansatz, to maintain the accuracy of DMRG with the increase of the system
size in the study of two dimensional systems, exponentially increased resources are required, which limits the applicability of DMRG to only narrow systems. In this work, we introduce a new ansatz in which DMRG is augmented with disentanglers to
encode {area-law-like entanglement entropy} ({entanglement entropy supported in the new ansatz scales as $l$ for a $l \times l$ system}). In the new method, the $O(D^3)$ low computational cost of DMRG is kept ({with an overhead of $O(d^4)$ and $d$ the dimension of the physical degree of freedom}). We
perform benchmark calculations with this approach on the two dimensional transverse Ising and Heisenberg models.
{This new ansatz extends the power of DMRG in the study of two-dimensional quantum systems.}
\end{abstract}

\maketitle


The study of exotic phases and the phase transitions between them in strongly correlated quantum many-body systems is one of the largest challenges in condensed matter physics \cite{bookc,Xiao:803748,marino_2017}. The main difficulty stems from the exponential growth of the Hilbert space dimension with the system size, which means brutal force
approach can only handle very small systems.
Analytic solution to quantum many-body systems is very rare \cite{Berlinsky2019,RevModPhys.76.643,PhysRevLett.20.1445,Qiao_2020,PhysRevLett.122.180401}. So most studies of these systems
rely on different types of numerical methods nowadays \cite{PhysRevB.26.5033,PhysRevB.56.11678,Huggins2022,PhysRevLett.69.2863,PhysRevB.48.10345,RevModPhys.77.259,SCHOLLWOCK201196,PhysRevX.5.041041,RevModPhys.93.045003}.

Density Matrix Renormalization Group (DMRG) \cite{PhysRevLett.69.2863,PhysRevB.48.10345,RevModPhys.77.259,SCHOLLWOCK201196,doi:10.1146/annurev-conmatphys-020911-125018} is the most successful numerical methods for the study of one dimensional (1D) quantum many-body systems \cite{PhysRevLett.109.056402,PhysRevLett.83.3297} in the past decades.
But it is known that the physics in two dimension (2D) is richer \cite{PhysRevLett.126.160604,Verzhbitskiy_2020,PhysRevB.103.L140101,DallaPiazza2015,Ludwig_2011,PhysRevLett.126.015301,PhysRevLett.128.146803,PhysRevLett.53.722,PhysRevLett.52.1583,PhysRevLett.49.957}.
The direct application of DMRG to 2D systems is not as successful as the study of 1D cases. It was
found that the required resource needs to increase exponentially with the system size in 2D if we want to maintain the accuracy \cite{PhysRevB.49.9214}.

It was realized later that the wave-functions obtained by DMRG are actually Matrix Product States (MPS) \cite{PhysRevLett.75.3537}.
and the success of DMRG lies in the fact that the entanglement encoded in the wave-function
satisfies the entropic area law \cite{PhysRevLett.94.060503,PhysRevLett.90.227902,PhysRevLett.71.666,RevModPhys.82.277} of 1D quantum systems. But MPS fail to capture the entropic area law for 2D systems which has to be remedied by exponentially large bond dimensions. To overcome this difficulty, MPS were generalized to high dimension in the perspective of Tensor Network States (TNS) \cite{Orus2019,Bridgeman_2017}. {There is also effort to modify the MPS ansatz for 2D systems \cite{PhysRevB.106.L081111}.}
In TNS, the wave-function of a quantum many-body system is represented as the contraction of connected tensors
with polynomial parameters in contrast to the exponentially large Hilbert space.
Different types of TNS were proposed in the past, like Projected Entangled Pair States (PEPS) \cite{PhysRevB.103.235155,PhysRevLett.125.210504,PhysRevX.9.031041,10.21468/SciPostPhys.5.5.047,PhysRevB.105.195140,RevModPhys.93.045003}, Tree Tensor Network (TTN) \cite{PhysRevLett.126.170603,PhysRevB.105.205102,10.21468/SciPostPhysLectNotes.8,Cataldi2021hilbertcurvevs}, Multiscale Entanglement Renormalization ansatz (MERA) \cite{PhysRevB.91.165129,PhysRevLett.101.110501,PhysRevLett.102.180406,PhysRevLett.99.220405}, and
projected entangled simplex states (PESS) \cite{PhysRevX.4.011025}.
It can be easily proven that 2D PEPS, MERA, and PESS can capture the entropic area law for 2D quantum systems \cite{PhysRevLett.96.220601,RevModPhys.82.277,PhysRevX.4.011025}
which makes them better wave-function ansatz for 2D systems.  
Progress in the understanding of exotic states in 2D has been made with these TNS-related methods in the past \cite{PhysRevB.103.235155,PhysRevLett.104.187203,PhysRevLett.112.147203,PhysRevLett.118.137202,Zheng17,LIU20221034}, but the high computational complexity
hampers the wide application of them in the study of 2D systems. The typical cost of PEPS is $O(D^{10})$ \cite{PhysRevB.90.064425} where $D$ is the bond dimension (we notice that many attempts have been made to low the computation cost of PEPS \cite{PhysRevB.96.045128, PhysRevB.98.235148, PhysRevB.103.235155,PhysRevB.102.125143}) and $O(D^{16})$ \cite{PhysRevLett.102.180406} for MERA, which makes the
calculation with large bond dimension infeasible and limit the power of these methods. For practical reason, because DMRG and MPS-based approaches are easy to be optimized and the computational cost is low, they are still widely used in the study of 2D (ladder or cylindrical) systems \cite{PhysRevLett.113.027201,doi:10.1126/science.1201080,PhysRevX.10.031016,PhysRevLett.127.097002,PhysRevLett.127.097003,doi:10.1073/pnas.2109978118} by pushing the bond dimension to very large numbers, even though they cannot capture the entropic area law for 2D systems intrinsically. 

\begin{figure}[t]
	 \includegraphics[width=90mm]{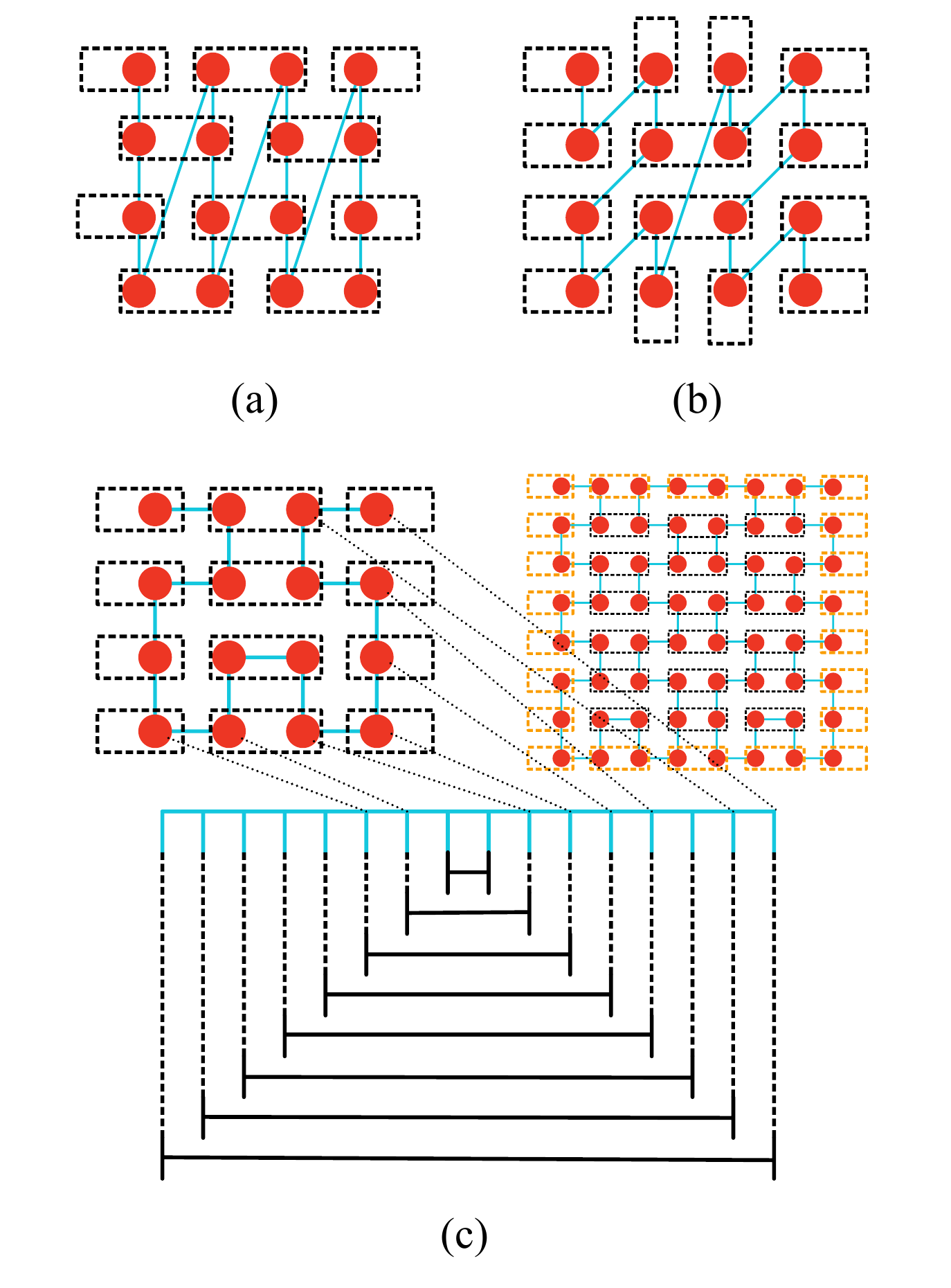}
	\caption{Different schemes to arrange a 2D lattice into a 1D one in DMRG for periodic boundary conditions (PBC). The disentanglers in corresponding
		FAMPS are denoted as dashed rectangles. (a) The commonly adopted scheme for a $4\times 4$ lattice. (b) A
		tree-type scheme for a $4\times 4$ lattice (more details can be found in \cite{PhysRevB.105.205102}). {(c) A new
		scheme used in this work for $4\times 4 $ and $8\times 8$ lattices and the corresponding one dimensional representation of the $4\times 4$ lattice where the dashed black lines indicate how disentanglers act on MPS. The disentanglers at the edges denoted as dashed orange rectangles need to be rearranged for different boundary
		conditions.  When augmented with disentanglers, the
		FAMPS encodes more entanglement in (b) and (c) than that in (a). See the main text for more discussion.}}
	\label{DMRG_arrangements}
\end{figure}

In this work, we introduce a new MPS or DMRG based ansatz dubbed as Fully-augmented Matrix Product State (FAMPS) by
generalizing the idea from augmented-TTN (aTTN) \cite{PhysRevLett.126.170603} and fully augmented-TTN (FATTN) \cite{PhysRevB.105.205102} to MPS.
{The entanglement entropy encoded in both TTN and MPS are bounded by $\log (D)$, but MPS has a lower cost than TTN.} In FAMPS, MPS is augmented with
disentanglers to increase the entanglement encoded in the wave-function. In the simplest scheme, where disentanglers are
placed directly in the physical layer and span only two sites, it can be proved that the entanglement entropy captured in FAMPS scales as $l\ln(d^2)$
with $l$ the measure of the cut with which the system is divided into two parts and $d$ is the dimension
of local Hilbert space. Most importantly, the low,
i.e., $O(D^3)$ computational cost in DMRG is maintained in FAMPS. 

\begin{figure}[t]
 \includegraphics[width=90mm]{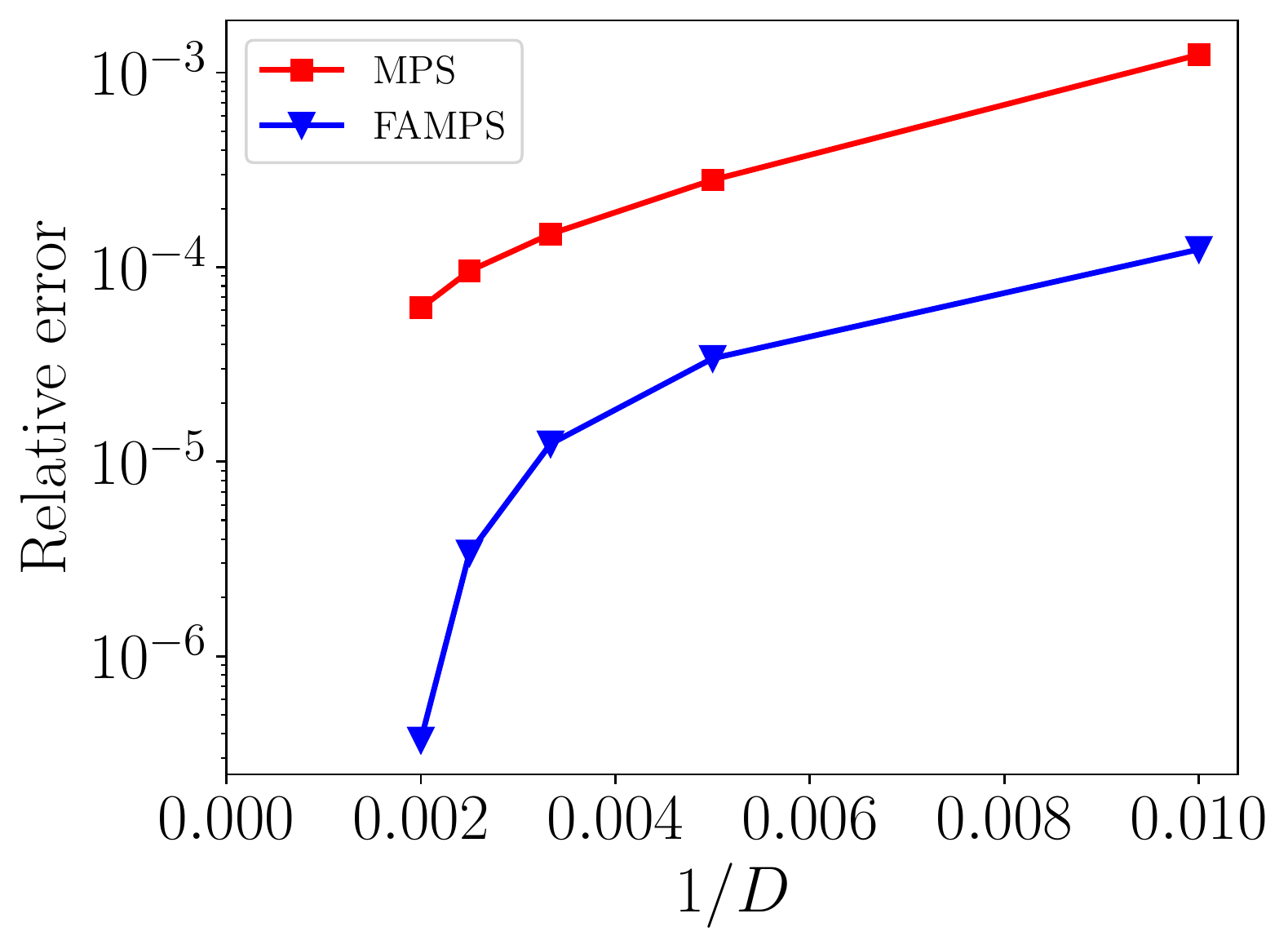}
	\caption{Relative error of the ground state energy for FAMPS and MPS of the transverse Ising model near the critical point ($\lambda = 3.05$) for a $8\times 8$ lattice with PBC.
	The scheme in Fig.~\ref{DMRG_arrangements} (c) is used.
		The reference energy is the {Quantum Monte Carlo (QMC) result calculated by the stochasitic series expansion (SSE) method with
		inverse temperature $\beta=240$ \cite{PhysRevE.66.046701,foot6}, which is -3.24163(1). Since our best energy of FAMPS $-3.2416288$ (with $D=500$)
		agrees with the reference energy within the error bar, the value of last point of FAMPS may be underestimated.} We can find that the improvement of FAMPS over MPS is about one order of magnitude.}
	\label{Ising_model}
\end{figure}

{\em Fully-Augmented Matrix Product States --} An MPS is defined as
\begin{equation}
 |\text{MPS} \rangle=\sum_{\{\sigma_i\}} \text{Tr}[A^{\sigma_1}A^{\sigma_2}A^{\sigma_3}\cdots A^{\sigma_n}]|\sigma_1 \sigma_2 \sigma_3\cdots \sigma_n\rangle
 \end{equation}
 where $A$ is a rank-3 tensor with one physical index $\sigma_i$ (with dimension $d$) and two auxiliary indices (with dimension $D$).

Disentanglers {are unitary matrices which transform the physical degrees of freedom of two sites.} They are common building blocks in TNS which can reduce the local entanglement in the studied system \cite{PhysRevLett.101.110501,PhysRevA.101.032310}. Following the strategy in \cite{PhysRevLett.126.170603,PhysRevB.105.205102}, we can place an additional disentangler layer on the physical layer of MPS to increase the entanglement in the wave-function as
\begin{equation}
|\text{FAMPS}\rangle=D(u)|\text{MPS}\rangle
\end{equation}where $D(u)=\prod_{m} u_m$ denotes the disentangler layer. {Fig.~\ref{DMRG_arrangements} (c) shows the diagrammatic representation of FAMPS, where we can find that FAMPS is a more entangled wave-function than MPS \cite{foot5}.}

As discussed in Ref. \cite{PhysRevB.105.205102}, there are two criteria or constraints when placing disentanglers. The first one is that there should be no two disentanglers sharing the same physical site. The other one is to place more disentanglers in
places where the entanglement the ansatz wave-function can host is small. The first criterion ensures a comparable computational complexity with its predecessor (with a factor of $d^4$ at the worst case)
The second criterion is to make the entanglement distribute as uniform as possible in the whole system. More specifically, we want to ensure the least entanglement in the wave-function for all different cuts as large as possible.

The special structure of FAMPS suggests that the optimization process of FAMPS can be divided into two steps. First, the disentanglers can be optimized using the standard Evenbly-Vidal algorithm \cite{PhysRevB.79.144108}. Second, the rest {of the} tensors can be directly optimized with the traditional MPS or DMRG optimization procedure
{by solving eigenvalue problems \cite{PhysRevLett.69.2863}. The optimization of disentanglers and MPS tensors are performed alternatively for many times
untill the energy is converged.} 
{The details can be found in the supplementary materials.}
{As discussed in the supplementary materials, the $O(D^3)$ cost of DMRG is kept in FAMPS with
an additional factor of $d^4$.}

The arrangement of a 2D lattice to a 1D one in DMRG was extensively studied in
the past \cite{PhysRevB.64.104414,Cataldi2021hilbertcurvevs}. In the DMRG calculation, different arrangements have small effect on the accuracy as will be
shown late. However, the way a 2D lattice is arranged in a 1D one is crucial when augmenting MPS with disentanglers, which determines the amount of entanglement the wave-function can encode. 
In Fig.~\ref{DMRG_arrangements}, we show three different ways to arrange the 2D lattice in a 1D one.
The scheme in Fig.~\ref{DMRG_arrangements} (a) is the most common used one in the literature. In this scheme, the entanglement is large if we cut the system horizontally, while only one bond is crossed if the system is cut vertically. 
This means an MPS in the scheme of Fig.~\ref{DMRG_arrangements} (a) can only encode an
entanglement entropy $S \le \ln(D)$.
To augment MPS with disentanglers based on the scheme in Fig.~\ref{DMRG_arrangements} (a),
we need to place all the disentanglers (the dashed rectangulars in Fig.~\ref{DMRG_arrangements} (a)) horizontally
according to the two criteria mentioned above. Because we can only place $L/2$ disentangles horizontally
in each column and the entanglement entropy contributed by each disentangler is maximally $\ln(d^2)$,
the maximum entanglement entropy the FAMPS in Fig.~\ref{DMRG_arrangements} (a) can encode {scales as $L\ln(d)$}.
 
We have more efficient way to place the disentanglers. In Fig.~\ref{DMRG_arrangements} (c), we introduce
another scheme to arrange the 2D lattice in a 1D one. The entanglement in the MPS in Fig.~\ref{DMRG_arrangements} (c)
is more uniformly distributed than that in Fig.~\ref{DMRG_arrangements} (a) \cite{foot1}. When augmented with disentanglers,
the maximum entanglement entropy encoded in the FAMPS in Fig.~\ref{DMRG_arrangements} (c) is $L\ln(d^2)$ which
is twice that in Fig.~\ref{DMRG_arrangements} (a) {and in the aTTN \cite{PhysRevLett.126.170603}}. Comparing to Fig.~\ref{DMRG_arrangements} (a), the number
of crossed bonds
with horizontal cut is reduced by a half, but even a small $D = d^4$ can support the $L\ln(d^2)$ entanglement entropy
for these cuts. Actually Fig.~\ref{DMRG_arrangements} (c) is not the only scheme to support the $L\ln(d^2)$ entanglement
entropy, another example is shown Fig.~\ref{DMRG_arrangements} (b), which is analyzed in Ref. \cite{PhysRevB.105.205102} based on TTN. 
{By placing the disentanglers appropriately, the arrangement following the Hilbert curve \cite{Cataldi2021hilbertcurvevs} can also support the $L\ln(d^2)$ entanglement
entropy.} In this work, we mainly focus on the scheme in Fig.~\ref{DMRG_arrangements} (c).

\begin{figure}[t]
     \includegraphics[width=80mm]{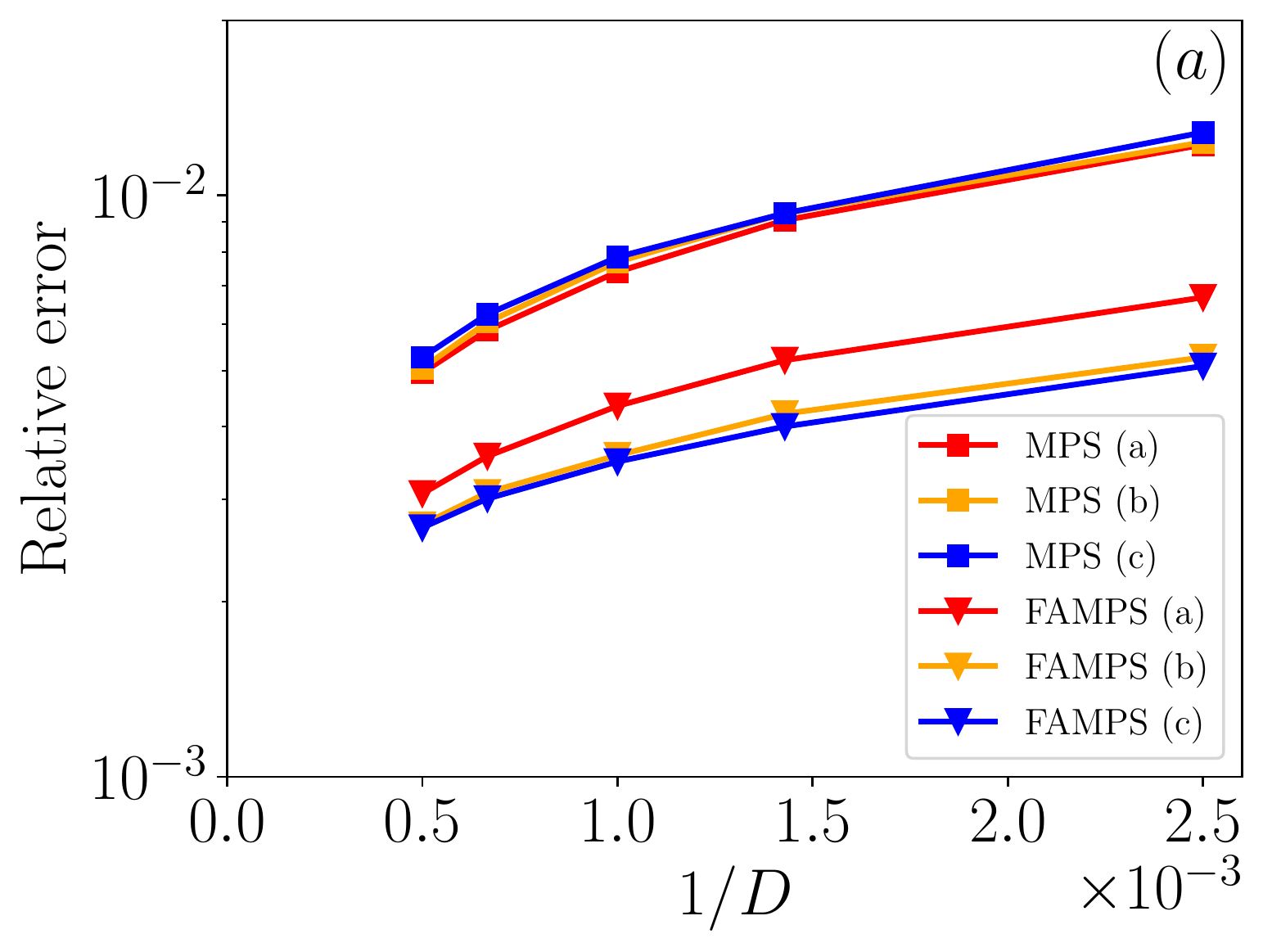}
    \includegraphics[width=80mm]{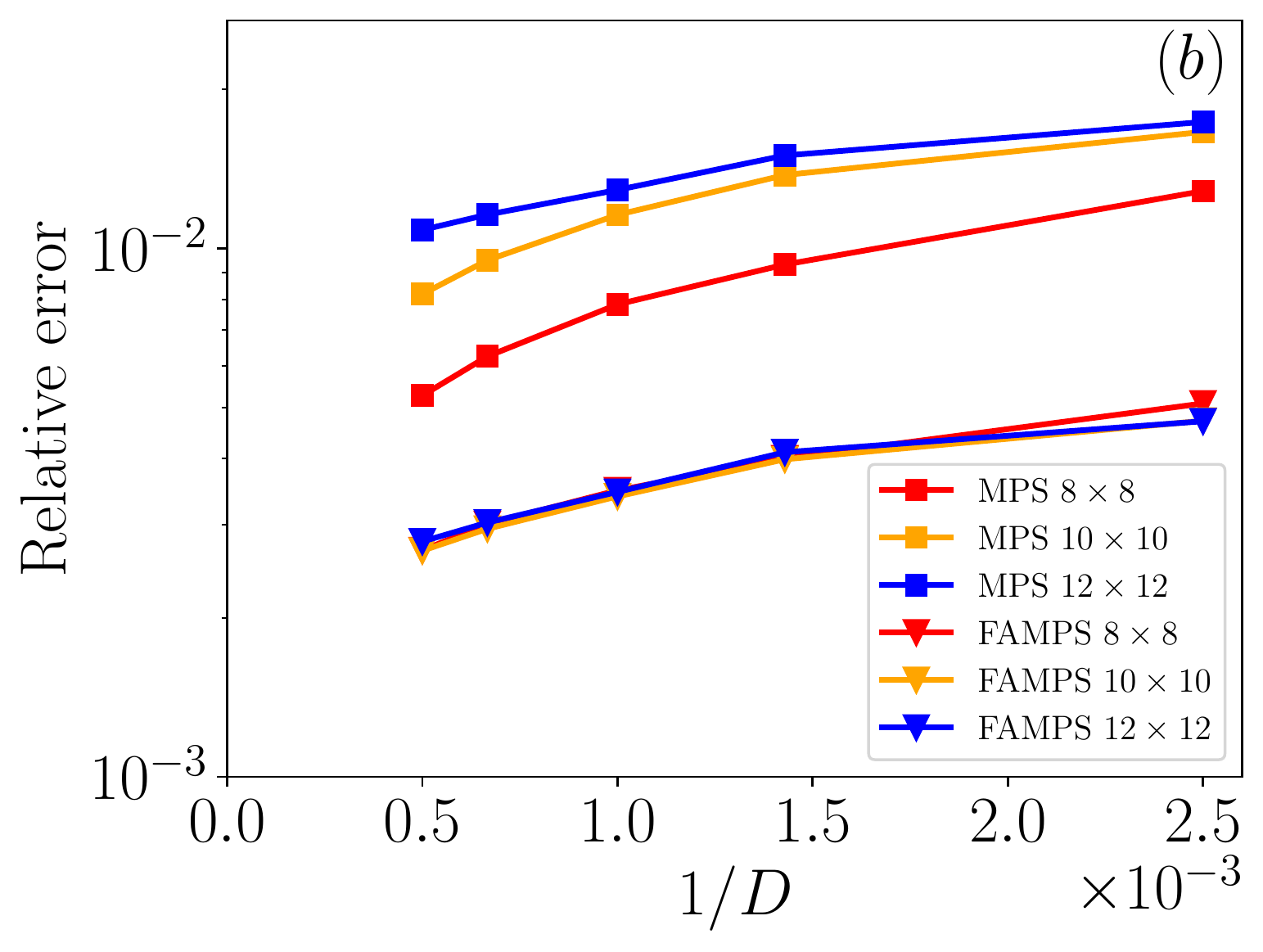}
    \caption{Relative error of the ground state energy for FAMPS and MPS of the Heisenberg model with PBC. The reference energies are the numerically exact Quantum Monte Carlo results \cite{PhysRevB.56.11678}.  (a) We compare the results of different schemes in Fig.~\ref{DMRG_arrangements}. ((a), (b), and (c) in the
    	legend denote the different schemes in Fig.~\ref{DMRG_arrangements} (a), (b), and (c) respectively). We can see that the different schemes have little differences in DMRG energy. But after augmented with disentanglers, the FAMPS in Fig.~\ref{DMRG_arrangements} (c) gives lower energy than that in Fig.~\ref{DMRG_arrangements} (a) as expected. (b) We compare the results of different lattice sizes using the scheme in  Fig.~\ref{DMRG_arrangements} (c). We can see that FAMPS accuracy is maintained with the increase of lattice size while MPS accuracy becomes worse as anticipated.}
    \label{Heisenberg_PBC}
\end{figure}

{\em Results on the transverse Ising model --}
We first test FAMPS in the two dimensional transverse Ising model.
The Hamiltonian of transverse Ising model is
\begin{equation}
H_{\text{Ising}}=-\sum_{\langle i,j \rangle}\sigma_{i}^{z}\sigma_{j}^{z}-\lambda\sum_{i}\sigma_{i}^{x}
\end{equation}
  where $\{\sigma_{i}^x,\sigma_{i}^{z}\}$ are Pauli matrices. In Fig.~\ref{Ising_model}, we show the relative error of ground state energy from FAMPS and MPS simulations
  with $\lambda = 3.05$ (close to the critical point of the model).
  The scheme in Fig.~\ref{DMRG_arrangements} (c) is used to map the 2D lattice into 1D.
The system is with size $8 \times 8$ and with PBC.
As shown in Fig.~\ref{Ising_model}, 
we can find that FAMPS is about one order of magnitude more accurate than MPS, which is similar to the improvement of FATTN over TTN in Ref. \cite{PhysRevB.105.205102}.

\begin{figure}[t]
	\includegraphics[width=80mm]{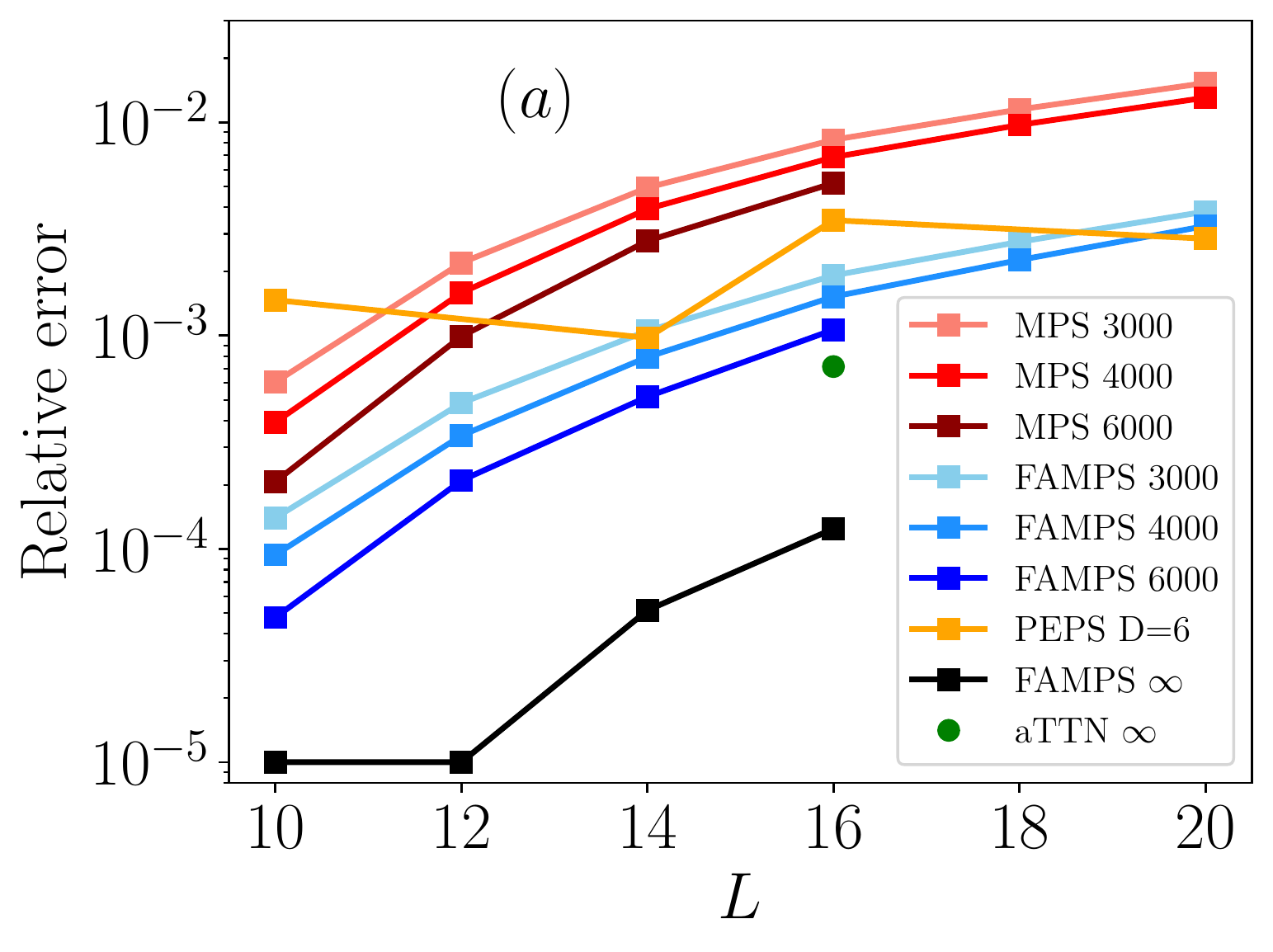}
	\includegraphics[width=40mm]{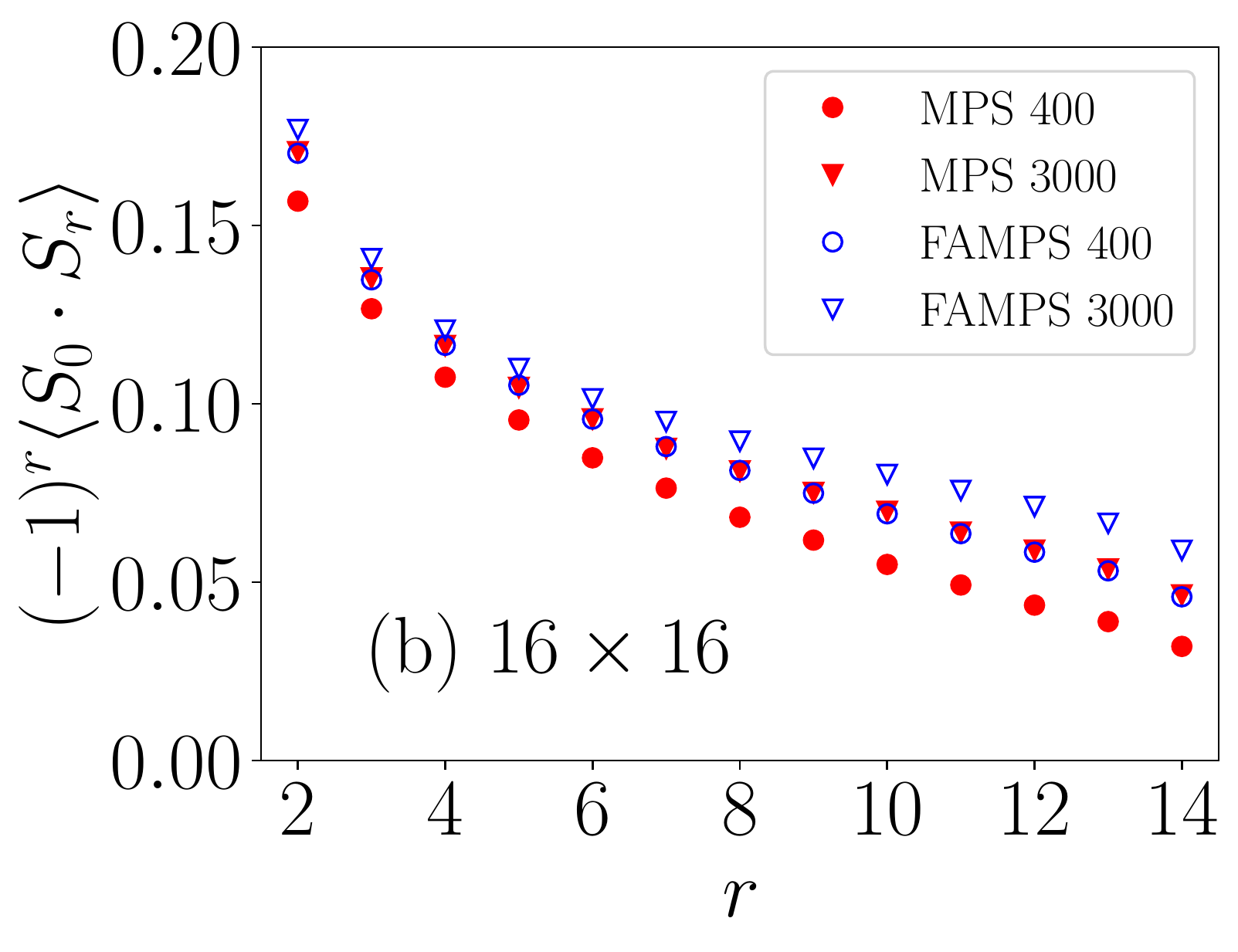}
	\includegraphics[width=40mm]{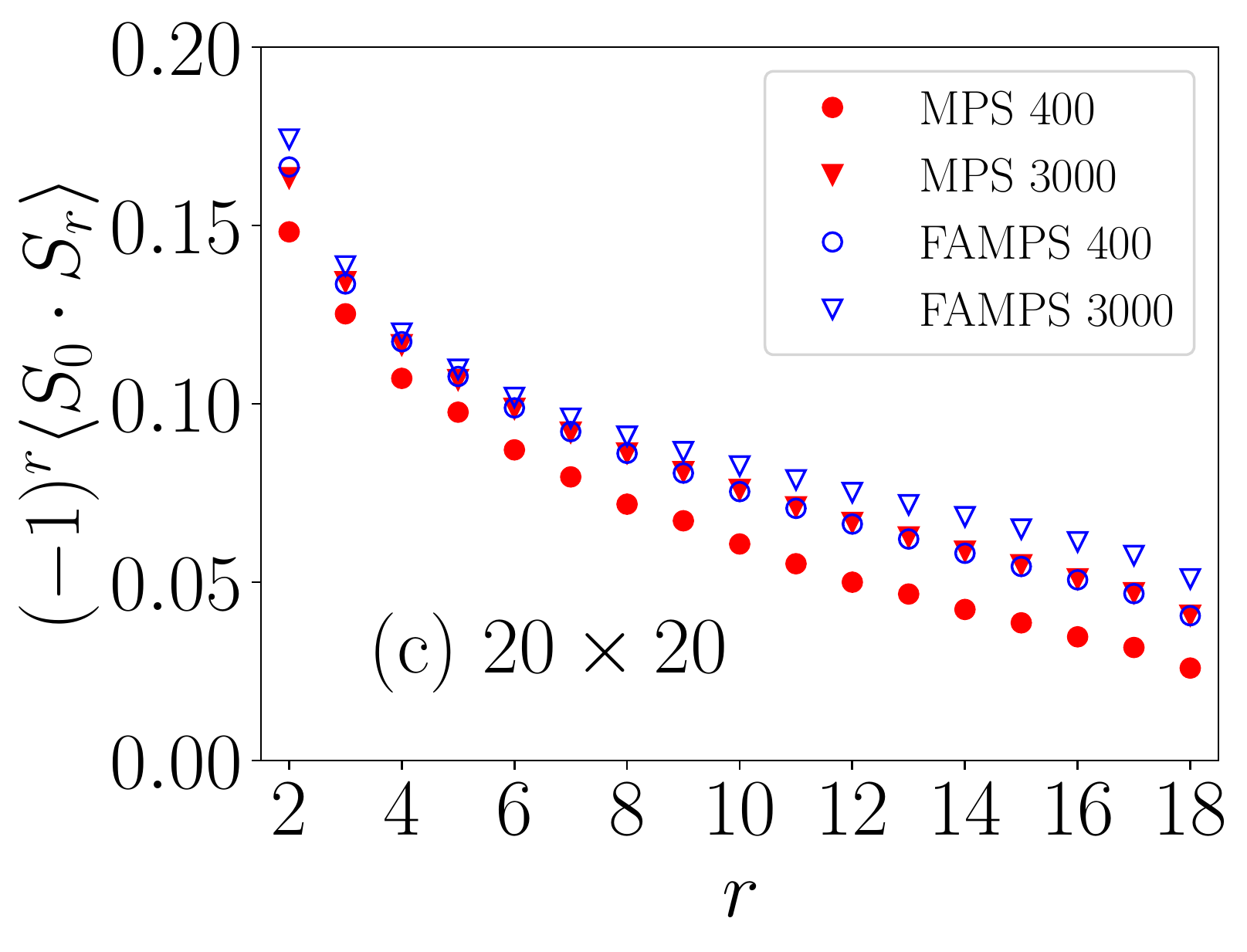}
   \caption{{ (a) Relative error of the ground state energy for FAMPS, MPS, aTTN \cite{PhysRevLett.126.170603}, and PEPS \cite{PhysRevB.90.064425}  of the Heisenberg model
   		on square lattice with size from $10 \times 10$ to $20 \times 20$ (squares and circles indicate open and periodic boundary conditions respectively). The reference energy is the numerically exact results obtained by Quantum Monte Carlo in \cite{PhysRevB.103.235155,PhysRevB.56.11678}. The black line is the extrapolated results for FAMPS at $D\rightarrow\infty$. For $L=10$ and $L=12$, because the extrapolated energies of FAMPS are exact within the error bar of Quantum Monte Carlo results, we set the relative errors of extrapolated FAMPS as $10^{-5}$. From the plot, we can find FAMPS is able to outperform PEPS for $L\le 20$. (b) and (c) show the corresponding spin correlations 
   		along the horizontal direction for $16\times 16$ and $20\times 20$ systems. FAMPS is able to give the same accuracy with a nearly one tenth bond dimension ($D = 400$) compared with MPS ($D = 3000$).}
   }
   \label{Heisenberg_OBC}
\end{figure}

{\em Results on the Heisenberg model --}
We also test FAMPS on the 2D Heisenberg model whose ground state is
more entangled than the transverse Ising model. The Hamiltonian of Heisenberg model is
\begin{equation}
 H_{\text{Heisenberg}}=\sum_{\langle i,j \rangle}S_{i} \cdot S_{j}
\end{equation}
where $S_{i}$ denote the spin of site $i$. We consider only nearest neighboring interactions. Here, we take
	advantage of the U(1) symmetry \cite{PhysRevB.83.115125} in our simulation. We take the numerically
exact Quantum Monte Carlo results \cite{PhysRevB.56.11678,PhysRevB.103.235155} as references.  First, we calculate the ground state energy for $8\times 8$ lattice with periodic boundary conditions to test FAMPS with different schemes to arrange the 2D lattice
in 1D. Fig.~\ref{Heisenberg_PBC} (a) shows the relative error of the ground state energy as a function of bond-dimension. We can find that the energies from MPS without disentangler layer have little differences for the three schemes in
Fig.~\ref{DMRG_arrangements}. This result agrees with the study in \cite{Cataldi2021hilbertcurvevs,PhysRevB.64.104414} which show different arrangements can lead to increased numerical precision but not significant for 2D systems. 
However, when augmented with disentanglers, FAMPS results in Fig.~\ref{Heisenberg_PBC} (a) show the differences in accuracy is significant for
different arrangement schemes. As expected, the FAMPS in Fig.~\ref{DMRG_arrangements} (c) is more accurate than
that in Fig.~\ref{DMRG_arrangements} (a) because the entanglement encoded in the wave-function ansatz in (c) is larger. For the Heisenberg model, the improvement in accuracy is about half an order of magnitude.

We then test FAMPS with larger sizes using the scheme in Fig.~\ref{DMRG_arrangements} (c)
In Fig.~\ref{Heisenberg_PBC} (b), we show the relative error of the ground state energy for lattice sizes $8\times 8$,
$10\times 10$, and $12\times 12$. The most important finding in these results is that the accuracy of FAMPS is maintained with the
increase of lattice size, despite the accuracy of MPS decreases with the increase of lattice size as expected.
This property makes FAMPS a promising method
in the study of two dimensional systems in the future.
 

{We also test FAMPS for Heisenberg model with open boundary conditions. We find that there exists a critical bond dimension after which the energies calculated by FAMPS and MPS are identical in the U(1) symmetry imposed calculation (see the plot in the supplementary materials). This problem is caused by the absence of free parameters of disentanglers. The disentangler for
	spin $1/2$ degree of freedom ($d=2$) has one free parameter under U(1) symmetry and no free parameter under SU(2) symmetry. When the bond dimension exceeds the critical value, the state in U(1) symmetry imposed MPS calculation nearly restores the SU(2) symmetry. Thus, the extra disentangler layer has no effect and FAMPS and MPS give the same results.
	The transverse Ising model has Z(2) symmetry with two free parameters in the disentangler, which is the reason why there is no critical bond dimension
	in Fig.~\ref{Ising_model}. The details of analyzing the number of free parameters in the disentanglers can be found in the Supplementary Materials. }
	
{To solve this problem, we block two sites into one to increase the number of free parameters of disentangler and impose SU(2) symmetry in the calculation. Fig.~\ref{Heisenberg_OBC} (a) shows the relative error of ground state energy calculated by MPS and FAMPS for different lattice sizes. {We can find that the eneries of FAMPS are lower than PEPS results with bond-dimension D=6 for $L\le 20$ in Fig.~\ref{Heisenberg_OBC} (a)}. When extrapolated to the infinite bond dimension limit for both FAMPS
	and aTTN,
FAMPS is also more accurate than aTTN. Fig.~\ref{Heisenberg_OBC} (b) and (c) show the corresponding spin correlations along the horizontal direction.
FAMPS gives the same results as MPS with a nearly one tenth bond dimension for the correlation function.}

{At last, we show preliminary results of FAMPS on the more challenging J1-J2 Heisenberg model. We focus on the most difficult point with $J_2/J1 = 0.5$. The comparison of ground state energy for a $8 \times 8$ system with PBC is listed in Table.~\ref{Energy}. We can find that the FAMPS energy with $D=12000$ is more accurate than previous DMRG energy with $D=32000$ and FAMPS with $D=24000$ gives the most accurate energy.}

\begin{table}[ht]
	\begin{tabular}{lll}
		\toprule
		Method& Energy per site\\
		\hline
		CNN & $-0.49639(2)$ \cite{PhysRevB.100.125124}\\
		RBM+PP &  $-0.498460(6)$ \cite{PhysRevX.11.031034}\\
		DMRG D=24000 &  $-0.497961$ \cite{PhysRevLett.113.027201}\\
		DMRG D=32000 & $-0.498175$ \cite{PhysRevLett.113.027201}\\
		VMC & $-0.49656(1)$ \cite{PhysRevB.88.060402}\\
		FAMPS D=12000 (this work)& $-0.498184$\\
		FAMPS D=24000 (this work)& $-0.498508$\\
		\hline
		
	\end{tabular}
	\caption{The comparison of the ground state energy of the J1-J2 Heisenberg model ($J_2/J_1 = 0.5$) on a $8 \times 8$ square lattice with periodic boundary conditions.}
	\label{Energy}  
\end{table}

{\em Conclusions --}
In this work, we propose a new ansatz, FAMPS, by augmenting MPS with disentanglers. FAMPS can encode {area-law-like entanglement entropy} for two dimensional quantum systems and at the same time keeps the low $O(D^3)$ computational cost ({with an overhead of $O(d^4)$}). We carry out benchmark
calculations of FAMPS on the 2D transverse Ising model near critical point and the 2D Heisenberg model. 
{FAMPS provides a useful approach to extend the power of DMRG for the study
of difficult 2D problems \cite{PhysRevLett.113.027201}}. FAMPS can be generalized to
encode more entanglement by placing multiple layers of disentangles to MPS. The {linearly scaled entanglement} encoded in FAMPS
can also benefit many other MPS based approaches, e.g., the simulation of time evolution or the dynamic properties \cite{PAECKEL2019167998,PhysRevLett.102.240603,PhysRevLett.93.076401}. {FAMPS is also applicable to any other lattices
by rearranging the studied system in a one dimension way.}

\begin{acknowledgments}
{The calculation in this work is carried out with Quimb \cite{gray2018quimb} and TensorKit \cite{foot7}. MQ acknowledges the
support from the National Key Research and Development Program of MOST of China (2022YFA1405400),
the National Natural Science Foundation of China (Grant
No. 12274290) and the sponsorship from Yangyang Development Fund.}
\end{acknowledgments}

\bibliography{FAMPS}

\begin{thebibliography}{85}
\expandafter\ifx\csname natexlab\endcsname\relax\def\natexlab#1{#1}\fi
\expandafter\ifx\csname bibnamefont\endcsname\relax
  \def\bibnamefont#1{#1}\fi
\expandafter\ifx\csname bibfnamefont\endcsname\relax
  \def\bibfnamefont#1{#1}\fi
\expandafter\ifx\csname citenamefont\endcsname\relax
  \def\citenamefont#1{#1}\fi
\expandafter\ifx\csname url\endcsname\relax
  \def\url#1{\texttt{#1}}\fi
\expandafter\ifx\csname urlprefix\endcsname\relax\def\urlprefix{URL }\fi
\providecommand{\bibinfo}[2]{#2}
\providecommand{\eprint}[2][]{\url{#2}}

\bibitem[{\citenamefont{Cabra et~al.}(2012)\citenamefont{Cabra, Honecker, and
  Pujol}}]{bookc}
\bibinfo{author}{\bibfnamefont{D.}~\bibnamefont{Cabra}},
  \bibinfo{author}{\bibfnamefont{A.}~\bibnamefont{Honecker}}, \bibnamefont{and}
  \bibinfo{author}{\bibfnamefont{P.}~\bibnamefont{Pujol}},
  \emph{\bibinfo{title}{Modern Theories of Many-Particle Systems in Condensed
  Matter Physics}}, vol. \bibinfo{volume}{843} (\bibinfo{year}{2012}), ISBN
  \bibinfo{isbn}{978-3-642-10448-0}.

\bibitem[{\citenamefont{Wen}(2007)}]{Xiao:803748}
\bibinfo{author}{\bibfnamefont{X.~G.} \bibnamefont{Wen}},
  \emph{\bibinfo{title}{{Quantum field theory of many-body systems: from the
  origin of sound to an origin of light and electrons}}}
  (\bibinfo{publisher}{Oxford University Press}, \bibinfo{address}{Oxford},
  \bibinfo{year}{2007}), \urlprefix\url{https://cds.cern.ch/record/803748}.

\bibitem[{\citenamefont{Marino}(2017)}]{marino_2017}
\bibinfo{author}{\bibfnamefont{E.~C.} \bibnamefont{Marino}},
  \emph{\bibinfo{title}{Quantum Field Theory Approach to Condensed Matter
  Physics}} (\bibinfo{publisher}{Cambridge University Press},
  \bibinfo{address}{Cambridge}, \bibinfo{year}{2017}).

\bibitem[{\citenamefont{Berlinsky and Harris}(2019)}]{Berlinsky2019}
\bibinfo{author}{\bibfnamefont{A.~J.} \bibnamefont{Berlinsky}}
  \bibnamefont{and} \bibinfo{author}{\bibfnamefont{A.~B.}
  \bibnamefont{Harris}}, \emph{\bibinfo{title}{The Ising Model: Exact
  Solutions}} (\bibinfo{publisher}{Springer International Publishing},
  \bibinfo{address}{Cham}, \bibinfo{year}{2019}), pp.
  \bibinfo{pages}{441--476}, ISBN \bibinfo{isbn}{978-3-030-28187-8},
  \urlprefix\url{https://doi.org/10.1007/978-3-030-28187-8_17}.

\bibitem[{\citenamefont{Dukelsky et~al.}(2004)\citenamefont{Dukelsky, Pittel,
  and Sierra}}]{RevModPhys.76.643}
\bibinfo{author}{\bibfnamefont{J.}~\bibnamefont{Dukelsky}},
  \bibinfo{author}{\bibfnamefont{S.}~\bibnamefont{Pittel}}, \bibnamefont{and}
  \bibinfo{author}{\bibfnamefont{G.}~\bibnamefont{Sierra}},
  \bibinfo{journal}{Rev. Mod. Phys.} \textbf{\bibinfo{volume}{76}},
  \bibinfo{pages}{643} (\bibinfo{year}{2004}),
  \urlprefix\url{https://link.aps.org/doi/10.1103/RevModPhys.76.643}.

\bibitem[{\citenamefont{Lieb and Wu}(1968)}]{PhysRevLett.20.1445}
\bibinfo{author}{\bibfnamefont{E.~H.} \bibnamefont{Lieb}} \bibnamefont{and}
  \bibinfo{author}{\bibfnamefont{F.~Y.} \bibnamefont{Wu}},
  \bibinfo{journal}{Phys. Rev. Lett.} \textbf{\bibinfo{volume}{20}},
  \bibinfo{pages}{1445} (\bibinfo{year}{1968}),
  \urlprefix\url{https://link.aps.org/doi/10.1103/PhysRevLett.20.1445}.

\bibitem[{\citenamefont{Qiao et~al.}(2020)\citenamefont{Qiao, Sun, Xin, Cao,
  and Yang}}]{Qiao_2020}
\bibinfo{author}{\bibfnamefont{Y.}~\bibnamefont{Qiao}},
  \bibinfo{author}{\bibfnamefont{P.}~\bibnamefont{Sun}},
  \bibinfo{author}{\bibfnamefont{Z.}~\bibnamefont{Xin}},
  \bibinfo{author}{\bibfnamefont{J.}~\bibnamefont{Cao}}, \bibnamefont{and}
  \bibinfo{author}{\bibfnamefont{W.-L.} \bibnamefont{Yang}},
  \bibinfo{journal}{Journal of Physics A: Mathematical and Theoretical}
  \textbf{\bibinfo{volume}{53}}, \bibinfo{pages}{075205}
  (\bibinfo{year}{2020}),
  \urlprefix\url{https://doi.org/10.1088/1751-8121/ab6a32}.

\bibitem[{\citenamefont{Zou et~al.}(2019)\citenamefont{Zou, Zhao, Guan, and
  Liu}}]{PhysRevLett.122.180401}
\bibinfo{author}{\bibfnamefont{H.}~\bibnamefont{Zou}},
  \bibinfo{author}{\bibfnamefont{E.}~\bibnamefont{Zhao}},
  \bibinfo{author}{\bibfnamefont{X.-W.} \bibnamefont{Guan}}, \bibnamefont{and}
  \bibinfo{author}{\bibfnamefont{W.~V.} \bibnamefont{Liu}},
  \bibinfo{journal}{Phys. Rev. Lett.} \textbf{\bibinfo{volume}{122}},
  \bibinfo{pages}{180401} (\bibinfo{year}{2019}),
  \urlprefix\url{https://link.aps.org/doi/10.1103/PhysRevLett.122.180401}.

\bibitem[{\citenamefont{Hirsch et~al.}(1982)\citenamefont{Hirsch, Sugar,
  Scalapino, and Blankenbecler}}]{PhysRevB.26.5033}
\bibinfo{author}{\bibfnamefont{J.~E.} \bibnamefont{Hirsch}},
  \bibinfo{author}{\bibfnamefont{R.~L.} \bibnamefont{Sugar}},
  \bibinfo{author}{\bibfnamefont{D.~J.} \bibnamefont{Scalapino}},
  \bibnamefont{and}
  \bibinfo{author}{\bibfnamefont{R.}~\bibnamefont{Blankenbecler}},
  \bibinfo{journal}{Phys. Rev. B} \textbf{\bibinfo{volume}{26}},
  \bibinfo{pages}{5033} (\bibinfo{year}{1982}),
  \urlprefix\url{https://link.aps.org/doi/10.1103/PhysRevB.26.5033}.

\bibitem[{\citenamefont{Sandvik}(1997)}]{PhysRevB.56.11678}
\bibinfo{author}{\bibfnamefont{A.~W.} \bibnamefont{Sandvik}},
  \bibinfo{journal}{Phys. Rev. B} \textbf{\bibinfo{volume}{56}},
  \bibinfo{pages}{11678} (\bibinfo{year}{1997}),
  \urlprefix\url{https://link.aps.org/doi/10.1103/PhysRevB.56.11678}.

\bibitem[{\citenamefont{Huggins et~al.}(2022)\citenamefont{Huggins, O'Gorman,
  Rubin, Reichman, Babbush, and Lee}}]{Huggins2022}
\bibinfo{author}{\bibfnamefont{W.~J.} \bibnamefont{Huggins}},
  \bibinfo{author}{\bibfnamefont{B.~A.} \bibnamefont{O'Gorman}},
  \bibinfo{author}{\bibfnamefont{N.~C.} \bibnamefont{Rubin}},
  \bibinfo{author}{\bibfnamefont{D.~R.} \bibnamefont{Reichman}},
  \bibinfo{author}{\bibfnamefont{R.}~\bibnamefont{Babbush}}, \bibnamefont{and}
  \bibinfo{author}{\bibfnamefont{J.}~\bibnamefont{Lee}},
  \bibinfo{journal}{Nature} \textbf{\bibinfo{volume}{603}},
  \bibinfo{pages}{416} (\bibinfo{year}{2022}), ISSN \bibinfo{issn}{1476-4687},
  \urlprefix\url{https://doi.org/10.1038/s41586-021-04351-z}.

\bibitem[{\citenamefont{White}(1992)}]{PhysRevLett.69.2863}
\bibinfo{author}{\bibfnamefont{S.~R.} \bibnamefont{White}},
  \bibinfo{journal}{Phys. Rev. Lett.} \textbf{\bibinfo{volume}{69}},
  \bibinfo{pages}{2863} (\bibinfo{year}{1992}),
  \urlprefix\url{https://link.aps.org/doi/10.1103/PhysRevLett.69.2863}.

\bibitem[{\citenamefont{White}(1993)}]{PhysRevB.48.10345}
\bibinfo{author}{\bibfnamefont{S.~R.} \bibnamefont{White}},
  \bibinfo{journal}{Phys. Rev. B} \textbf{\bibinfo{volume}{48}},
  \bibinfo{pages}{10345} (\bibinfo{year}{1993}),
  \urlprefix\url{https://link.aps.org/doi/10.1103/PhysRevB.48.10345}.

\bibitem[{\citenamefont{Schollw\"ock}(2005)}]{RevModPhys.77.259}
\bibinfo{author}{\bibfnamefont{U.}~\bibnamefont{Schollw\"ock}},
  \bibinfo{journal}{Rev. Mod. Phys.} \textbf{\bibinfo{volume}{77}},
  \bibinfo{pages}{259} (\bibinfo{year}{2005}),
  \urlprefix\url{https://link.aps.org/doi/10.1103/RevModPhys.77.259}.

\bibitem[{\citenamefont{Schollw\"ock}(2011)}]{SCHOLLWOCK201196}
\bibinfo{author}{\bibfnamefont{U.}~\bibnamefont{Schollw\"ock}},
  \bibinfo{journal}{Annals of Physics} \textbf{\bibinfo{volume}{326}},
  \bibinfo{pages}{96} (\bibinfo{year}{2011}), ISSN \bibinfo{issn}{0003-4916},
  \bibinfo{note}{january 2011 Special Issue},
  \urlprefix\url{https://www.sciencedirect.com/science/article/pii/S0003491610001752}.

\bibitem[{\citenamefont{LeBlanc et~al.}(2015)\citenamefont{LeBlanc, Antipov,
  Becca, Bulik, Chan, Chung, Deng, Ferrero, Henderson, Jim\'enez-Hoyos
  et~al.}}]{PhysRevX.5.041041}
\bibinfo{author}{\bibfnamefont{J.~P.~F.} \bibnamefont{LeBlanc}},
  \bibinfo{author}{\bibfnamefont{A.~E.} \bibnamefont{Antipov}},
  \bibinfo{author}{\bibfnamefont{F.}~\bibnamefont{Becca}},
  \bibinfo{author}{\bibfnamefont{I.~W.} \bibnamefont{Bulik}},
  \bibinfo{author}{\bibfnamefont{G.~K.-L.} \bibnamefont{Chan}},
  \bibinfo{author}{\bibfnamefont{C.-M.} \bibnamefont{Chung}},
  \bibinfo{author}{\bibfnamefont{Y.}~\bibnamefont{Deng}},
  \bibinfo{author}{\bibfnamefont{M.}~\bibnamefont{Ferrero}},
  \bibinfo{author}{\bibfnamefont{T.~M.} \bibnamefont{Henderson}},
  \bibinfo{author}{\bibfnamefont{C.~A.} \bibnamefont{Jim\'enez-Hoyos}},
  \bibnamefont{et~al.} (\bibinfo{collaboration}{Simons Collaboration on the
  Many-Electron Problem}), \bibinfo{journal}{Phys. Rev. X}
  \textbf{\bibinfo{volume}{5}}, \bibinfo{pages}{041041} (\bibinfo{year}{2015}),
  \urlprefix\url{https://link.aps.org/doi/10.1103/PhysRevX.5.041041}.

\bibitem[{\citenamefont{Cirac et~al.}(2021)\citenamefont{Cirac,
  P\'erez-Garc\'{\i}a, Schuch, and Verstraete}}]{RevModPhys.93.045003}
\bibinfo{author}{\bibfnamefont{J.~I.} \bibnamefont{Cirac}},
  \bibinfo{author}{\bibfnamefont{D.}~\bibnamefont{P\'erez-Garc\'{\i}a}},
  \bibinfo{author}{\bibfnamefont{N.}~\bibnamefont{Schuch}}, \bibnamefont{and}
  \bibinfo{author}{\bibfnamefont{F.}~\bibnamefont{Verstraete}},
  \bibinfo{journal}{Rev. Mod. Phys.} \textbf{\bibinfo{volume}{93}},
  \bibinfo{pages}{045003} (\bibinfo{year}{2021}),
  \urlprefix\url{https://link.aps.org/doi/10.1103/RevModPhys.93.045003}.

\bibitem[{\citenamefont{Stoudenmire and
  White}(2012)}]{doi:10.1146/annurev-conmatphys-020911-125018}
\bibinfo{author}{\bibfnamefont{E.}~\bibnamefont{Stoudenmire}} \bibnamefont{and}
  \bibinfo{author}{\bibfnamefont{S.~R.} \bibnamefont{White}},
  \bibinfo{journal}{Annual Review of Condensed Matter Physics}
  \textbf{\bibinfo{volume}{3}}, \bibinfo{pages}{111} (\bibinfo{year}{2012}),
  \eprint{https://doi.org/10.1146/annurev-conmatphys-020911-125018},
  \urlprefix\url{https://doi.org/10.1146/annurev-conmatphys-020911-125018}.

\bibitem[{\citenamefont{Stoudenmire et~al.}(2012)\citenamefont{Stoudenmire,
  Wagner, White, and Burke}}]{PhysRevLett.109.056402}
\bibinfo{author}{\bibfnamefont{E.~M.} \bibnamefont{Stoudenmire}},
  \bibinfo{author}{\bibfnamefont{L.~O.} \bibnamefont{Wagner}},
  \bibinfo{author}{\bibfnamefont{S.~R.} \bibnamefont{White}}, \bibnamefont{and}
  \bibinfo{author}{\bibfnamefont{K.}~\bibnamefont{Burke}},
  \bibinfo{journal}{Phys. Rev. Lett.} \textbf{\bibinfo{volume}{109}},
  \bibinfo{pages}{056402} (\bibinfo{year}{2012}),
  \urlprefix\url{https://link.aps.org/doi/10.1103/PhysRevLett.109.056402}.

\bibitem[{\citenamefont{Hida}(1999)}]{PhysRevLett.83.3297}
\bibinfo{author}{\bibfnamefont{K.}~\bibnamefont{Hida}}, \bibinfo{journal}{Phys.
  Rev. Lett.} \textbf{\bibinfo{volume}{83}}, \bibinfo{pages}{3297}
  (\bibinfo{year}{1999}),
  \urlprefix\url{https://link.aps.org/doi/10.1103/PhysRevLett.83.3297}.

\bibitem[{\citenamefont{Nakano et~al.}(2021)\citenamefont{Nakano, Minami, and
  Sasa}}]{PhysRevLett.126.160604}
\bibinfo{author}{\bibfnamefont{H.}~\bibnamefont{Nakano}},
  \bibinfo{author}{\bibfnamefont{Y.}~\bibnamefont{Minami}}, \bibnamefont{and}
  \bibinfo{author}{\bibfnamefont{S.-i.} \bibnamefont{Sasa}},
  \bibinfo{journal}{Phys. Rev. Lett.} \textbf{\bibinfo{volume}{126}},
  \bibinfo{pages}{160604} (\bibinfo{year}{2021}),
  \urlprefix\url{https://link.aps.org/doi/10.1103/PhysRevLett.126.160604}.

\bibitem[{\citenamefont{Verzhbitskiy et~al.}(2020)\citenamefont{Verzhbitskiy,
  Voiry, Chhowalla, and Eda}}]{Verzhbitskiy_2020}
\bibinfo{author}{\bibfnamefont{I.~A.} \bibnamefont{Verzhbitskiy}},
  \bibinfo{author}{\bibfnamefont{D.}~\bibnamefont{Voiry}},
  \bibinfo{author}{\bibfnamefont{M.}~\bibnamefont{Chhowalla}},
  \bibnamefont{and} \bibinfo{author}{\bibfnamefont{G.}~\bibnamefont{Eda}},
  \bibinfo{journal}{2D Materials} \textbf{\bibinfo{volume}{7}},
  \bibinfo{pages}{035013} (\bibinfo{year}{2020}),
  \urlprefix\url{https://doi.org/10.1088/2053-1583/ab8690}.

\bibitem[{\citenamefont{Astrakharchik et~al.}(2021)\citenamefont{Astrakharchik,
  Kurbakov, Sychev, Fedorov, and Lozovik}}]{PhysRevB.103.L140101}
\bibinfo{author}{\bibfnamefont{G.~E.} \bibnamefont{Astrakharchik}},
  \bibinfo{author}{\bibfnamefont{I.~L.} \bibnamefont{Kurbakov}},
  \bibinfo{author}{\bibfnamefont{D.~V.} \bibnamefont{Sychev}},
  \bibinfo{author}{\bibfnamefont{A.~K.} \bibnamefont{Fedorov}},
  \bibnamefont{and} \bibinfo{author}{\bibfnamefont{Y.~E.}
  \bibnamefont{Lozovik}}, \bibinfo{journal}{Phys. Rev. B}
  \textbf{\bibinfo{volume}{103}}, \bibinfo{pages}{L140101}
  (\bibinfo{year}{2021}),
  \urlprefix\url{https://link.aps.org/doi/10.1103/PhysRevB.103.L140101}.

\bibitem[{\citenamefont{Dalla~Piazza et~al.}(2015)\citenamefont{Dalla~Piazza,
  Mourigal, Christensen, Nilsen, Tregenna-Piggott, Perring, Enderle, McMorrow,
  Ivanov, and R{\o}nnow}}]{DallaPiazza2015}
\bibinfo{author}{\bibfnamefont{B.}~\bibnamefont{Dalla~Piazza}},
  \bibinfo{author}{\bibfnamefont{M.}~\bibnamefont{Mourigal}},
  \bibinfo{author}{\bibfnamefont{N.~B.} \bibnamefont{Christensen}},
  \bibinfo{author}{\bibfnamefont{G.~J.} \bibnamefont{Nilsen}},
  \bibinfo{author}{\bibfnamefont{P.}~\bibnamefont{Tregenna-Piggott}},
  \bibinfo{author}{\bibfnamefont{T.~G.} \bibnamefont{Perring}},
  \bibinfo{author}{\bibfnamefont{M.}~\bibnamefont{Enderle}},
  \bibinfo{author}{\bibfnamefont{D.~F.} \bibnamefont{McMorrow}},
  \bibinfo{author}{\bibfnamefont{D.~A.} \bibnamefont{Ivanov}},
  \bibnamefont{and} \bibinfo{author}{\bibfnamefont{H.~M.}
  \bibnamefont{R{\o}nnow}}, \bibinfo{journal}{Nature Physics}
  \textbf{\bibinfo{volume}{11}}, \bibinfo{pages}{62} (\bibinfo{year}{2015}),
  ISSN \bibinfo{issn}{1745-2481},
  \urlprefix\url{https://doi.org/10.1038/nphys3172}.

\bibitem[{\citenamefont{Ludwig et~al.}(2011)\citenamefont{Ludwig, Poilblanc,
  Trebst, and Troyer}}]{Ludwig_2011}
\bibinfo{author}{\bibfnamefont{A.~W.~W.} \bibnamefont{Ludwig}},
  \bibinfo{author}{\bibfnamefont{D.}~\bibnamefont{Poilblanc}},
  \bibinfo{author}{\bibfnamefont{S.}~\bibnamefont{Trebst}}, \bibnamefont{and}
  \bibinfo{author}{\bibfnamefont{M.}~\bibnamefont{Troyer}},
  \bibinfo{journal}{New Journal of Physics} \textbf{\bibinfo{volume}{13}},
  \bibinfo{pages}{045014} (\bibinfo{year}{2011}),
  \urlprefix\url{https://doi.org/10.1088/1367-2630/13/4/045014}.

\bibitem[{\citenamefont{Brooks et~al.}(2021)\citenamefont{Brooks, Lemeshko,
  Lundholm, and Yakaboylu}}]{PhysRevLett.126.015301}
\bibinfo{author}{\bibfnamefont{M.}~\bibnamefont{Brooks}},
  \bibinfo{author}{\bibfnamefont{M.}~\bibnamefont{Lemeshko}},
  \bibinfo{author}{\bibfnamefont{D.}~\bibnamefont{Lundholm}}, \bibnamefont{and}
  \bibinfo{author}{\bibfnamefont{E.}~\bibnamefont{Yakaboylu}},
  \bibinfo{journal}{Phys. Rev. Lett.} \textbf{\bibinfo{volume}{126}},
  \bibinfo{pages}{015301} (\bibinfo{year}{2021}),
  \urlprefix\url{https://link.aps.org/doi/10.1103/PhysRevLett.126.015301}.

\bibitem[{\citenamefont{Han et~al.}(2022)\citenamefont{Han, Iftikhar, Kleeorin,
  Anthore, Pierre, Meir, Mitchell, and Sela}}]{PhysRevLett.128.146803}
\bibinfo{author}{\bibfnamefont{C.}~\bibnamefont{Han}},
  \bibinfo{author}{\bibfnamefont{Z.}~\bibnamefont{Iftikhar}},
  \bibinfo{author}{\bibfnamefont{Y.}~\bibnamefont{Kleeorin}},
  \bibinfo{author}{\bibfnamefont{A.}~\bibnamefont{Anthore}},
  \bibinfo{author}{\bibfnamefont{F.}~\bibnamefont{Pierre}},
  \bibinfo{author}{\bibfnamefont{Y.}~\bibnamefont{Meir}},
  \bibinfo{author}{\bibfnamefont{A.~K.} \bibnamefont{Mitchell}},
  \bibnamefont{and} \bibinfo{author}{\bibfnamefont{E.}~\bibnamefont{Sela}},
  \bibinfo{journal}{Phys. Rev. Lett.} \textbf{\bibinfo{volume}{128}},
  \bibinfo{pages}{146803} (\bibinfo{year}{2022}),
  \urlprefix\url{https://link.aps.org/doi/10.1103/PhysRevLett.128.146803}.

\bibitem[{\citenamefont{Arovas et~al.}(1984)\citenamefont{Arovas, Schrieffer,
  and Wilczek}}]{PhysRevLett.53.722}
\bibinfo{author}{\bibfnamefont{D.}~\bibnamefont{Arovas}},
  \bibinfo{author}{\bibfnamefont{J.~R.} \bibnamefont{Schrieffer}},
  \bibnamefont{and} \bibinfo{author}{\bibfnamefont{F.}~\bibnamefont{Wilczek}},
  \bibinfo{journal}{Phys. Rev. Lett.} \textbf{\bibinfo{volume}{53}},
  \bibinfo{pages}{722} (\bibinfo{year}{1984}),
  \urlprefix\url{https://link.aps.org/doi/10.1103/PhysRevLett.53.722}.

\bibitem[{\citenamefont{Halperin}(1984)}]{PhysRevLett.52.1583}
\bibinfo{author}{\bibfnamefont{B.~I.} \bibnamefont{Halperin}},
  \bibinfo{journal}{Phys. Rev. Lett.} \textbf{\bibinfo{volume}{52}},
  \bibinfo{pages}{1583} (\bibinfo{year}{1984}),
  \urlprefix\url{https://link.aps.org/doi/10.1103/PhysRevLett.52.1583}.

\bibitem[{\citenamefont{Wilczek}(1982)}]{PhysRevLett.49.957}
\bibinfo{author}{\bibfnamefont{F.}~\bibnamefont{Wilczek}},
  \bibinfo{journal}{Phys. Rev. Lett.} \textbf{\bibinfo{volume}{49}},
  \bibinfo{pages}{957} (\bibinfo{year}{1982}),
  \urlprefix\url{https://link.aps.org/doi/10.1103/PhysRevLett.49.957}.

\bibitem[{\citenamefont{Liang and Pang}(1994)}]{PhysRevB.49.9214}
\bibinfo{author}{\bibfnamefont{S.}~\bibnamefont{Liang}} \bibnamefont{and}
  \bibinfo{author}{\bibfnamefont{H.}~\bibnamefont{Pang}},
  \bibinfo{journal}{Phys. Rev. B} \textbf{\bibinfo{volume}{49}},
  \bibinfo{pages}{9214} (\bibinfo{year}{1994}),
  \urlprefix\url{https://link.aps.org/doi/10.1103/PhysRevB.49.9214}.

\bibitem[{\citenamefont{\"Ostlund and Rommer}(1995)}]{PhysRevLett.75.3537}
\bibinfo{author}{\bibfnamefont{S.}~\bibnamefont{\"Ostlund}} \bibnamefont{and}
  \bibinfo{author}{\bibfnamefont{S.}~\bibnamefont{Rommer}},
  \bibinfo{journal}{Phys. Rev. Lett.} \textbf{\bibinfo{volume}{75}},
  \bibinfo{pages}{3537} (\bibinfo{year}{1995}),
  \urlprefix\url{https://link.aps.org/doi/10.1103/PhysRevLett.75.3537}.

\bibitem[{\citenamefont{Plenio et~al.}(2005)\citenamefont{Plenio, Eisert,
  Drei\ss{}ig, and Cramer}}]{PhysRevLett.94.060503}
\bibinfo{author}{\bibfnamefont{M.~B.} \bibnamefont{Plenio}},
  \bibinfo{author}{\bibfnamefont{J.}~\bibnamefont{Eisert}},
  \bibinfo{author}{\bibfnamefont{J.}~\bibnamefont{Drei\ss{}ig}},
  \bibnamefont{and} \bibinfo{author}{\bibfnamefont{M.}~\bibnamefont{Cramer}},
  \bibinfo{journal}{Phys. Rev. Lett.} \textbf{\bibinfo{volume}{94}},
  \bibinfo{pages}{060503} (\bibinfo{year}{2005}),
  \urlprefix\url{https://link.aps.org/doi/10.1103/PhysRevLett.94.060503}.

\bibitem[{\citenamefont{Vidal et~al.}(2003)\citenamefont{Vidal, Latorre, Rico,
  and Kitaev}}]{PhysRevLett.90.227902}
\bibinfo{author}{\bibfnamefont{G.}~\bibnamefont{Vidal}},
  \bibinfo{author}{\bibfnamefont{J.~I.} \bibnamefont{Latorre}},
  \bibinfo{author}{\bibfnamefont{E.}~\bibnamefont{Rico}}, \bibnamefont{and}
  \bibinfo{author}{\bibfnamefont{A.}~\bibnamefont{Kitaev}},
  \bibinfo{journal}{Phys. Rev. Lett.} \textbf{\bibinfo{volume}{90}},
  \bibinfo{pages}{227902} (\bibinfo{year}{2003}),
  \urlprefix\url{https://link.aps.org/doi/10.1103/PhysRevLett.90.227902}.

\bibitem[{\citenamefont{Srednicki}(1993)}]{PhysRevLett.71.666}
\bibinfo{author}{\bibfnamefont{M.}~\bibnamefont{Srednicki}},
  \bibinfo{journal}{Phys. Rev. Lett.} \textbf{\bibinfo{volume}{71}},
  \bibinfo{pages}{666} (\bibinfo{year}{1993}),
  \urlprefix\url{https://link.aps.org/doi/10.1103/PhysRevLett.71.666}.

\bibitem[{\citenamefont{Eisert et~al.}(2010)\citenamefont{Eisert, Cramer, and
  Plenio}}]{RevModPhys.82.277}
\bibinfo{author}{\bibfnamefont{J.}~\bibnamefont{Eisert}},
  \bibinfo{author}{\bibfnamefont{M.}~\bibnamefont{Cramer}}, \bibnamefont{and}
  \bibinfo{author}{\bibfnamefont{M.~B.} \bibnamefont{Plenio}},
  \bibinfo{journal}{Rev. Mod. Phys.} \textbf{\bibinfo{volume}{82}},
  \bibinfo{pages}{277} (\bibinfo{year}{2010}),
  \urlprefix\url{https://link.aps.org/doi/10.1103/RevModPhys.82.277}.

\bibitem[{\citenamefont{Or{\'u}s}(2019)}]{Orus2019}
\bibinfo{author}{\bibfnamefont{R.}~\bibnamefont{Or{\'u}s}},
  \bibinfo{journal}{Nature Reviews Physics} \textbf{\bibinfo{volume}{1}},
  \bibinfo{pages}{538} (\bibinfo{year}{2019}), ISSN \bibinfo{issn}{2522-5820},
  \urlprefix\url{https://doi.org/10.1038/s42254-019-0086-7}.

\bibitem[{\citenamefont{Bridgeman and Chubb}(2017)}]{Bridgeman_2017}
\bibinfo{author}{\bibfnamefont{J.~C.} \bibnamefont{Bridgeman}}
  \bibnamefont{and} \bibinfo{author}{\bibfnamefont{C.~T.} \bibnamefont{Chubb}},
  \bibinfo{journal}{Journal of Physics A: Mathematical and Theoretical}
  \textbf{\bibinfo{volume}{50}}, \bibinfo{pages}{223001}
  (\bibinfo{year}{2017}),
  \urlprefix\url{https://doi.org/10.1088/1751-8121/aa6dc3}.

\bibitem[{\citenamefont{Lami et~al.}(2022)\citenamefont{Lami, Carleo, and
  Collura}}]{PhysRevB.106.L081111}
\bibinfo{author}{\bibfnamefont{G.}~\bibnamefont{Lami}},
  \bibinfo{author}{\bibfnamefont{G.}~\bibnamefont{Carleo}}, \bibnamefont{and}
  \bibinfo{author}{\bibfnamefont{M.}~\bibnamefont{Collura}},
  \bibinfo{journal}{Phys. Rev. B} \textbf{\bibinfo{volume}{106}},
  \bibinfo{pages}{L081111} (\bibinfo{year}{2022}),
  \urlprefix\url{https://link.aps.org/doi/10.1103/PhysRevB.106.L081111}.

\bibitem[{\citenamefont{Liu et~al.}(2021)\citenamefont{Liu, Huang, Gong, and
  Gu}}]{PhysRevB.103.235155}
\bibinfo{author}{\bibfnamefont{W.-Y.} \bibnamefont{Liu}},
  \bibinfo{author}{\bibfnamefont{Y.-Z.} \bibnamefont{Huang}},
  \bibinfo{author}{\bibfnamefont{S.-S.} \bibnamefont{Gong}}, \bibnamefont{and}
  \bibinfo{author}{\bibfnamefont{Z.-C.} \bibnamefont{Gu}},
  \bibinfo{journal}{Phys. Rev. B} \textbf{\bibinfo{volume}{103}},
  \bibinfo{pages}{235155} (\bibinfo{year}{2021}),
  \urlprefix\url{https://link.aps.org/doi/10.1103/PhysRevB.103.235155}.

\bibitem[{\citenamefont{Scarpa et~al.}(2020)\citenamefont{Scarpa, Moln\'ar, Ge,
  Garc\'{\i}a-Ripoll, Schuch, P\'erez-Garc\'{\i}a, and
  Iblisdir}}]{PhysRevLett.125.210504}
\bibinfo{author}{\bibfnamefont{G.}~\bibnamefont{Scarpa}},
  \bibinfo{author}{\bibfnamefont{A.}~\bibnamefont{Moln\'ar}},
  \bibinfo{author}{\bibfnamefont{Y.}~\bibnamefont{Ge}},
  \bibinfo{author}{\bibfnamefont{J.~J.} \bibnamefont{Garc\'{\i}a-Ripoll}},
  \bibinfo{author}{\bibfnamefont{N.}~\bibnamefont{Schuch}},
  \bibinfo{author}{\bibfnamefont{D.}~\bibnamefont{P\'erez-Garc\'{\i}a}},
  \bibnamefont{and} \bibinfo{author}{\bibfnamefont{S.}~\bibnamefont{Iblisdir}},
  \bibinfo{journal}{Phys. Rev. Lett.} \textbf{\bibinfo{volume}{125}},
  \bibinfo{pages}{210504} (\bibinfo{year}{2020}),
  \urlprefix\url{https://link.aps.org/doi/10.1103/PhysRevLett.125.210504}.

\bibitem[{\citenamefont{Liao et~al.}(2019)\citenamefont{Liao, Liu, Wang, and
  Xiang}}]{PhysRevX.9.031041}
\bibinfo{author}{\bibfnamefont{H.-J.} \bibnamefont{Liao}},
  \bibinfo{author}{\bibfnamefont{J.-G.} \bibnamefont{Liu}},
  \bibinfo{author}{\bibfnamefont{L.}~\bibnamefont{Wang}}, \bibnamefont{and}
  \bibinfo{author}{\bibfnamefont{T.}~\bibnamefont{Xiang}},
  \bibinfo{journal}{Phys. Rev. X} \textbf{\bibinfo{volume}{9}},
  \bibinfo{pages}{031041} (\bibinfo{year}{2019}),
  \urlprefix\url{https://link.aps.org/doi/10.1103/PhysRevX.9.031041}.

\bibitem[{\citenamefont{Hubig}(2018)}]{10.21468/SciPostPhys.5.5.047}
\bibinfo{author}{\bibfnamefont{C.}~\bibnamefont{Hubig}},
  \bibinfo{journal}{SciPost Phys.} \textbf{\bibinfo{volume}{5}},
  \bibinfo{pages}{47} (\bibinfo{year}{2018}),
  \urlprefix\url{https://scipost.org/10.21468/SciPostPhys.5.5.047}.

\bibitem[{\citenamefont{Vanderstraeten
  et~al.}(2022)\citenamefont{Vanderstraeten, Burgelman, Ponsioen, Van~Damme,
  Vanhecke, Corboz, Haegeman, and Verstraete}}]{PhysRevB.105.195140}
\bibinfo{author}{\bibfnamefont{L.}~\bibnamefont{Vanderstraeten}},
  \bibinfo{author}{\bibfnamefont{L.}~\bibnamefont{Burgelman}},
  \bibinfo{author}{\bibfnamefont{B.}~\bibnamefont{Ponsioen}},
  \bibinfo{author}{\bibfnamefont{M.}~\bibnamefont{Van~Damme}},
  \bibinfo{author}{\bibfnamefont{B.}~\bibnamefont{Vanhecke}},
  \bibinfo{author}{\bibfnamefont{P.}~\bibnamefont{Corboz}},
  \bibinfo{author}{\bibfnamefont{J.}~\bibnamefont{Haegeman}}, \bibnamefont{and}
  \bibinfo{author}{\bibfnamefont{F.}~\bibnamefont{Verstraete}},
  \bibinfo{journal}{Phys. Rev. B} \textbf{\bibinfo{volume}{105}},
  \bibinfo{pages}{195140} (\bibinfo{year}{2022}),
  \urlprefix\url{https://link.aps.org/doi/10.1103/PhysRevB.105.195140}.

\bibitem[{\citenamefont{Felser et~al.}(2021)\citenamefont{Felser, Notarnicola,
  and Montangero}}]{PhysRevLett.126.170603}
\bibinfo{author}{\bibfnamefont{T.}~\bibnamefont{Felser}},
  \bibinfo{author}{\bibfnamefont{S.}~\bibnamefont{Notarnicola}},
  \bibnamefont{and}
  \bibinfo{author}{\bibfnamefont{S.}~\bibnamefont{Montangero}},
  \bibinfo{journal}{Phys. Rev. Lett.} \textbf{\bibinfo{volume}{126}},
  \bibinfo{pages}{170603} (\bibinfo{year}{2021}),
  \urlprefix\url{https://link.aps.org/doi/10.1103/PhysRevLett.126.170603}.

\bibitem[{\citenamefont{Qian and Qin}(2022)}]{PhysRevB.105.205102}
\bibinfo{author}{\bibfnamefont{X.}~\bibnamefont{Qian}} \bibnamefont{and}
  \bibinfo{author}{\bibfnamefont{M.}~\bibnamefont{Qin}},
  \bibinfo{journal}{Phys. Rev. B} \textbf{\bibinfo{volume}{105}},
  \bibinfo{pages}{205102} (\bibinfo{year}{2022}),
  \urlprefix\url{https://link.aps.org/doi/10.1103/PhysRevB.105.205102}.

\bibitem[{\citenamefont{Silvi et~al.}(2019)\citenamefont{Silvi, Tschirsich,
  Gerster, J\"unemann, Jaschke, Rizzi, and
  Montangero}}]{10.21468/SciPostPhysLectNotes.8}
\bibinfo{author}{\bibfnamefont{P.}~\bibnamefont{Silvi}},
  \bibinfo{author}{\bibfnamefont{F.}~\bibnamefont{Tschirsich}},
  \bibinfo{author}{\bibfnamefont{M.}~\bibnamefont{Gerster}},
  \bibinfo{author}{\bibfnamefont{J.}~\bibnamefont{J\"unemann}},
  \bibinfo{author}{\bibfnamefont{D.}~\bibnamefont{Jaschke}},
  \bibinfo{author}{\bibfnamefont{M.}~\bibnamefont{Rizzi}}, \bibnamefont{and}
  \bibinfo{author}{\bibfnamefont{S.}~\bibnamefont{Montangero}},
  \bibinfo{journal}{SciPost Phys. Lect. Notes} p.~\bibinfo{pages}{8}
  (\bibinfo{year}{2019}),
  \urlprefix\url{https://scipost.org/10.21468/SciPostPhysLectNotes.8}.

\bibitem[{\citenamefont{Cataldi et~al.}(2021)\citenamefont{Cataldi, Abedi,
  Magnifico, Notarnicola, Pozza, Giovannetti, and
  Montangero}}]{Cataldi2021hilbertcurvevs}
\bibinfo{author}{\bibfnamefont{G.}~\bibnamefont{Cataldi}},
  \bibinfo{author}{\bibfnamefont{A.}~\bibnamefont{Abedi}},
  \bibinfo{author}{\bibfnamefont{G.}~\bibnamefont{Magnifico}},
  \bibinfo{author}{\bibfnamefont{S.}~\bibnamefont{Notarnicola}},
  \bibinfo{author}{\bibfnamefont{N.~D.} \bibnamefont{Pozza}},
  \bibinfo{author}{\bibfnamefont{V.}~\bibnamefont{Giovannetti}},
  \bibnamefont{and}
  \bibinfo{author}{\bibfnamefont{S.}~\bibnamefont{Montangero}},
  \bibinfo{journal}{{Quantum}} \textbf{\bibinfo{volume}{5}},
  \bibinfo{pages}{556} (\bibinfo{year}{2021}), ISSN \bibinfo{issn}{2521-327X},
  \urlprefix\url{https://doi.org/10.22331/q-2021-09-29-556}.

\bibitem[{\citenamefont{Bridgeman et~al.}(2015)\citenamefont{Bridgeman,
  O'Brien, Bartlett, and Doherty}}]{PhysRevB.91.165129}
\bibinfo{author}{\bibfnamefont{J.~C.} \bibnamefont{Bridgeman}},
  \bibinfo{author}{\bibfnamefont{A.}~\bibnamefont{O'Brien}},
  \bibinfo{author}{\bibfnamefont{S.~D.} \bibnamefont{Bartlett}},
  \bibnamefont{and} \bibinfo{author}{\bibfnamefont{A.~C.}
  \bibnamefont{Doherty}}, \bibinfo{journal}{Phys. Rev. B}
  \textbf{\bibinfo{volume}{91}}, \bibinfo{pages}{165129}
  (\bibinfo{year}{2015}),
  \urlprefix\url{https://link.aps.org/doi/10.1103/PhysRevB.91.165129}.

\bibitem[{\citenamefont{Vidal}(2008)}]{PhysRevLett.101.110501}
\bibinfo{author}{\bibfnamefont{G.}~\bibnamefont{Vidal}},
  \bibinfo{journal}{Phys. Rev. Lett.} \textbf{\bibinfo{volume}{101}},
  \bibinfo{pages}{110501} (\bibinfo{year}{2008}),
  \urlprefix\url{https://link.aps.org/doi/10.1103/PhysRevLett.101.110501}.

\bibitem[{\citenamefont{Evenbly and
  Vidal}(2009{\natexlab{a}})}]{PhysRevLett.102.180406}
\bibinfo{author}{\bibfnamefont{G.}~\bibnamefont{Evenbly}} \bibnamefont{and}
  \bibinfo{author}{\bibfnamefont{G.}~\bibnamefont{Vidal}},
  \bibinfo{journal}{Phys. Rev. Lett.} \textbf{\bibinfo{volume}{102}},
  \bibinfo{pages}{180406} (\bibinfo{year}{2009}{\natexlab{a}}),
  \urlprefix\url{https://link.aps.org/doi/10.1103/PhysRevLett.102.180406}.

\bibitem[{\citenamefont{Vidal}(2007)}]{PhysRevLett.99.220405}
\bibinfo{author}{\bibfnamefont{G.}~\bibnamefont{Vidal}},
  \bibinfo{journal}{Phys. Rev. Lett.} \textbf{\bibinfo{volume}{99}},
  \bibinfo{pages}{220405} (\bibinfo{year}{2007}),
  \urlprefix\url{https://link.aps.org/doi/10.1103/PhysRevLett.99.220405}.

\bibitem[{\citenamefont{Xie et~al.}(2014)\citenamefont{Xie, Chen, Yu, Kong,
  Normand, and Xiang}}]{PhysRevX.4.011025}
\bibinfo{author}{\bibfnamefont{Z.~Y.} \bibnamefont{Xie}},
  \bibinfo{author}{\bibfnamefont{J.}~\bibnamefont{Chen}},
  \bibinfo{author}{\bibfnamefont{J.~F.} \bibnamefont{Yu}},
  \bibinfo{author}{\bibfnamefont{X.}~\bibnamefont{Kong}},
  \bibinfo{author}{\bibfnamefont{B.}~\bibnamefont{Normand}}, \bibnamefont{and}
  \bibinfo{author}{\bibfnamefont{T.}~\bibnamefont{Xiang}},
  \bibinfo{journal}{Phys. Rev. X} \textbf{\bibinfo{volume}{4}},
  \bibinfo{pages}{011025} (\bibinfo{year}{2014}),
  \urlprefix\url{https://link.aps.org/doi/10.1103/PhysRevX.4.011025}.

\bibitem[{\citenamefont{Verstraete et~al.}(2006)\citenamefont{Verstraete, Wolf,
  Perez-Garcia, and Cirac}}]{PhysRevLett.96.220601}
\bibinfo{author}{\bibfnamefont{F.}~\bibnamefont{Verstraete}},
  \bibinfo{author}{\bibfnamefont{M.~M.} \bibnamefont{Wolf}},
  \bibinfo{author}{\bibfnamefont{D.}~\bibnamefont{Perez-Garcia}},
  \bibnamefont{and} \bibinfo{author}{\bibfnamefont{J.~I.} \bibnamefont{Cirac}},
  \bibinfo{journal}{Phys. Rev. Lett.} \textbf{\bibinfo{volume}{96}},
  \bibinfo{pages}{220601} (\bibinfo{year}{2006}),
  \urlprefix\url{https://link.aps.org/doi/10.1103/PhysRevLett.96.220601}.

\bibitem[{\citenamefont{Evenbly and Vidal}(2010)}]{PhysRevLett.104.187203}
\bibinfo{author}{\bibfnamefont{G.}~\bibnamefont{Evenbly}} \bibnamefont{and}
  \bibinfo{author}{\bibfnamefont{G.}~\bibnamefont{Vidal}},
  \bibinfo{journal}{Phys. Rev. Lett.} \textbf{\bibinfo{volume}{104}},
  \bibinfo{pages}{187203} (\bibinfo{year}{2010}),
  \urlprefix\url{https://link.aps.org/doi/10.1103/PhysRevLett.104.187203}.

\bibitem[{\citenamefont{Corboz and Mila}(2014)}]{PhysRevLett.112.147203}
\bibinfo{author}{\bibfnamefont{P.}~\bibnamefont{Corboz}} \bibnamefont{and}
  \bibinfo{author}{\bibfnamefont{F.}~\bibnamefont{Mila}},
  \bibinfo{journal}{Phys. Rev. Lett.} \textbf{\bibinfo{volume}{112}},
  \bibinfo{pages}{147203} (\bibinfo{year}{2014}),
  \urlprefix\url{https://link.aps.org/doi/10.1103/PhysRevLett.112.147203}.

\bibitem[{\citenamefont{Liao et~al.}(2017)\citenamefont{Liao, Xie, Chen, Liu,
  Xie, Huang, Normand, and Xiang}}]{PhysRevLett.118.137202}
\bibinfo{author}{\bibfnamefont{H.~J.} \bibnamefont{Liao}},
  \bibinfo{author}{\bibfnamefont{Z.~Y.} \bibnamefont{Xie}},
  \bibinfo{author}{\bibfnamefont{J.}~\bibnamefont{Chen}},
  \bibinfo{author}{\bibfnamefont{Z.~Y.} \bibnamefont{Liu}},
  \bibinfo{author}{\bibfnamefont{H.~D.} \bibnamefont{Xie}},
  \bibinfo{author}{\bibfnamefont{R.~Z.} \bibnamefont{Huang}},
  \bibinfo{author}{\bibfnamefont{B.}~\bibnamefont{Normand}}, \bibnamefont{and}
  \bibinfo{author}{\bibfnamefont{T.}~\bibnamefont{Xiang}},
  \bibinfo{journal}{Phys. Rev. Lett.} \textbf{\bibinfo{volume}{118}},
  \bibinfo{pages}{137202} (\bibinfo{year}{2017}),
  \urlprefix\url{https://link.aps.org/doi/10.1103/PhysRevLett.118.137202}.

\bibitem[{\citenamefont{Zheng et~al.}(2017)\citenamefont{Zheng, Chung, Corboz,
  Ehlers, Qin, Noack, Shi, White, Zhang, and Chan}}]{Zheng17}
\bibinfo{author}{\bibfnamefont{B.-X.} \bibnamefont{Zheng}},
  \bibinfo{author}{\bibfnamefont{C.-M.} \bibnamefont{Chung}},
  \bibinfo{author}{\bibfnamefont{P.}~\bibnamefont{Corboz}},
  \bibinfo{author}{\bibfnamefont{G.}~\bibnamefont{Ehlers}},
  \bibinfo{author}{\bibfnamefont{M.-P.} \bibnamefont{Qin}},
  \bibinfo{author}{\bibfnamefont{R.~M.} \bibnamefont{Noack}},
  \bibinfo{author}{\bibfnamefont{H.}~\bibnamefont{Shi}},
  \bibinfo{author}{\bibfnamefont{S.~R.} \bibnamefont{White}},
  \bibinfo{author}{\bibfnamefont{S.}~\bibnamefont{Zhang}}, \bibnamefont{and}
  \bibinfo{author}{\bibfnamefont{G.~K.-L.} \bibnamefont{Chan}},
  \bibinfo{journal}{Science} \textbf{\bibinfo{volume}{358}},
  \bibinfo{pages}{1155} (\bibinfo{year}{2017}), ISSN \bibinfo{issn}{0036-8075,
  1095-9203},
  \urlprefix\url{http://science.sciencemag.org/content/358/6367/1155}.

\bibitem[{\citenamefont{Liu et~al.}(2022)\citenamefont{Liu, Gong, Li,
  Poilblanc, Chen, and Gu}}]{LIU20221034}
\bibinfo{author}{\bibfnamefont{W.-Y.} \bibnamefont{Liu}},
  \bibinfo{author}{\bibfnamefont{S.-S.} \bibnamefont{Gong}},
  \bibinfo{author}{\bibfnamefont{Y.-B.} \bibnamefont{Li}},
  \bibinfo{author}{\bibfnamefont{D.}~\bibnamefont{Poilblanc}},
  \bibinfo{author}{\bibfnamefont{W.-Q.} \bibnamefont{Chen}}, \bibnamefont{and}
  \bibinfo{author}{\bibfnamefont{Z.-C.} \bibnamefont{Gu}},
  \bibinfo{journal}{Science Bulletin} \textbf{\bibinfo{volume}{67}},
  \bibinfo{pages}{1034} (\bibinfo{year}{2022}), ISSN \bibinfo{issn}{2095-9273},
  \urlprefix\url{https://www.sciencedirect.com/science/article/pii/S2095927322001001}.

\bibitem[{\citenamefont{Lubasch et~al.}(2014)\citenamefont{Lubasch, Cirac, and
  Ba\~nuls}}]{PhysRevB.90.064425}
\bibinfo{author}{\bibfnamefont{M.}~\bibnamefont{Lubasch}},
  \bibinfo{author}{\bibfnamefont{J.~I.} \bibnamefont{Cirac}}, \bibnamefont{and}
  \bibinfo{author}{\bibfnamefont{M.-C.} \bibnamefont{Ba\~nuls}},
  \bibinfo{journal}{Phys. Rev. B} \textbf{\bibinfo{volume}{90}},
  \bibinfo{pages}{064425} (\bibinfo{year}{2014}),
  \urlprefix\url{https://link.aps.org/doi/10.1103/PhysRevB.90.064425}.

\bibitem[{\citenamefont{Xie et~al.}(2017)\citenamefont{Xie, Liao, Huang, Xie,
  Chen, Liu, and Xiang}}]{PhysRevB.96.045128}
\bibinfo{author}{\bibfnamefont{Z.~Y.} \bibnamefont{Xie}},
  \bibinfo{author}{\bibfnamefont{H.~J.} \bibnamefont{Liao}},
  \bibinfo{author}{\bibfnamefont{R.~Z.} \bibnamefont{Huang}},
  \bibinfo{author}{\bibfnamefont{H.~D.} \bibnamefont{Xie}},
  \bibinfo{author}{\bibfnamefont{J.}~\bibnamefont{Chen}},
  \bibinfo{author}{\bibfnamefont{Z.~Y.} \bibnamefont{Liu}}, \bibnamefont{and}
  \bibinfo{author}{\bibfnamefont{T.}~\bibnamefont{Xiang}},
  \bibinfo{journal}{Phys. Rev. B} \textbf{\bibinfo{volume}{96}},
  \bibinfo{pages}{045128} (\bibinfo{year}{2017}),
  \urlprefix\url{https://link.aps.org/doi/10.1103/PhysRevB.96.045128}.

\bibitem[{\citenamefont{Fishman et~al.}(2018)\citenamefont{Fishman,
  Vanderstraeten, Zauner-Stauber, Haegeman, and
  Verstraete}}]{PhysRevB.98.235148}
\bibinfo{author}{\bibfnamefont{M.~T.} \bibnamefont{Fishman}},
  \bibinfo{author}{\bibfnamefont{L.}~\bibnamefont{Vanderstraeten}},
  \bibinfo{author}{\bibfnamefont{V.}~\bibnamefont{Zauner-Stauber}},
  \bibinfo{author}{\bibfnamefont{J.}~\bibnamefont{Haegeman}}, \bibnamefont{and}
  \bibinfo{author}{\bibfnamefont{F.}~\bibnamefont{Verstraete}},
  \bibinfo{journal}{Phys. Rev. B} \textbf{\bibinfo{volume}{98}},
  \bibinfo{pages}{235148} (\bibinfo{year}{2018}),
  \urlprefix\url{https://link.aps.org/doi/10.1103/PhysRevB.98.235148}.

\bibitem[{\citenamefont{Qin}(2020)}]{PhysRevB.102.125143}
\bibinfo{author}{\bibfnamefont{M.}~\bibnamefont{Qin}}, \bibinfo{journal}{Phys.
  Rev. B} \textbf{\bibinfo{volume}{102}}, \bibinfo{pages}{125143}
  (\bibinfo{year}{2020}),
  \urlprefix\url{https://link.aps.org/doi/10.1103/PhysRevB.102.125143}.

\bibitem[{\citenamefont{Gong et~al.}(2014)\citenamefont{Gong, Zhu, Sheng,
  Motrunich, and Fisher}}]{PhysRevLett.113.027201}
\bibinfo{author}{\bibfnamefont{S.-S.} \bibnamefont{Gong}},
  \bibinfo{author}{\bibfnamefont{W.}~\bibnamefont{Zhu}},
  \bibinfo{author}{\bibfnamefont{D.~N.} \bibnamefont{Sheng}},
  \bibinfo{author}{\bibfnamefont{O.~I.} \bibnamefont{Motrunich}},
  \bibnamefont{and} \bibinfo{author}{\bibfnamefont{M.~P.~A.}
  \bibnamefont{Fisher}}, \bibinfo{journal}{Phys. Rev. Lett.}
  \textbf{\bibinfo{volume}{113}}, \bibinfo{pages}{027201}
  (\bibinfo{year}{2014}),
  \urlprefix\url{https://link.aps.org/doi/10.1103/PhysRevLett.113.027201}.

\bibitem[{\citenamefont{Yan et~al.}(2011)\citenamefont{Yan, Huse, and
  White}}]{doi:10.1126/science.1201080}
\bibinfo{author}{\bibfnamefont{S.}~\bibnamefont{Yan}},
  \bibinfo{author}{\bibfnamefont{D.~A.} \bibnamefont{Huse}}, \bibnamefont{and}
  \bibinfo{author}{\bibfnamefont{S.~R.} \bibnamefont{White}},
  \bibinfo{journal}{Science} \textbf{\bibinfo{volume}{332}},
  \bibinfo{pages}{1173} (\bibinfo{year}{2011}),
  \eprint{https://www.science.org/doi/pdf/10.1126/science.1201080},
  \urlprefix\url{https://www.science.org/doi/abs/10.1126/science.1201080}.

\bibitem[{\citenamefont{Qin et~al.}(2020)\citenamefont{Qin, Chung, Shi, Vitali,
  Hubig, Schollw\"ock, White, and Zhang}}]{PhysRevX.10.031016}
\bibinfo{author}{\bibfnamefont{M.}~\bibnamefont{Qin}},
  \bibinfo{author}{\bibfnamefont{C.-M.} \bibnamefont{Chung}},
  \bibinfo{author}{\bibfnamefont{H.}~\bibnamefont{Shi}},
  \bibinfo{author}{\bibfnamefont{E.}~\bibnamefont{Vitali}},
  \bibinfo{author}{\bibfnamefont{C.}~\bibnamefont{Hubig}},
  \bibinfo{author}{\bibfnamefont{U.}~\bibnamefont{Schollw\"ock}},
  \bibinfo{author}{\bibfnamefont{S.~R.} \bibnamefont{White}}, \bibnamefont{and}
  \bibinfo{author}{\bibfnamefont{S.}~\bibnamefont{Zhang}}
  (\bibinfo{collaboration}{Simons Collaboration on the Many-Electron Problem}),
  \bibinfo{journal}{Phys. Rev. X} \textbf{\bibinfo{volume}{10}},
  \bibinfo{pages}{031016} (\bibinfo{year}{2020}),
  \urlprefix\url{https://link.aps.org/doi/10.1103/PhysRevX.10.031016}.

\bibitem[{\citenamefont{Jiang and Kivelson}(2021)}]{PhysRevLett.127.097002}
\bibinfo{author}{\bibfnamefont{H.-C.} \bibnamefont{Jiang}} \bibnamefont{and}
  \bibinfo{author}{\bibfnamefont{S.~A.} \bibnamefont{Kivelson}},
  \bibinfo{journal}{Phys. Rev. Lett.} \textbf{\bibinfo{volume}{127}},
  \bibinfo{pages}{097002} (\bibinfo{year}{2021}),
  \urlprefix\url{https://link.aps.org/doi/10.1103/PhysRevLett.127.097002}.

\bibitem[{\citenamefont{Gong et~al.}(2021)\citenamefont{Gong, Zhu, and
  Sheng}}]{PhysRevLett.127.097003}
\bibinfo{author}{\bibfnamefont{S.}~\bibnamefont{Gong}},
  \bibinfo{author}{\bibfnamefont{W.}~\bibnamefont{Zhu}}, \bibnamefont{and}
  \bibinfo{author}{\bibfnamefont{D.~N.} \bibnamefont{Sheng}},
  \bibinfo{journal}{Phys. Rev. Lett.} \textbf{\bibinfo{volume}{127}},
  \bibinfo{pages}{097003} (\bibinfo{year}{2021}),
  \urlprefix\url{https://link.aps.org/doi/10.1103/PhysRevLett.127.097003}.

\bibitem[{\citenamefont{Jiang et~al.}(2021)\citenamefont{Jiang, Scalapino, and
  White}}]{doi:10.1073/pnas.2109978118}
\bibinfo{author}{\bibfnamefont{S.}~\bibnamefont{Jiang}},
  \bibinfo{author}{\bibfnamefont{D.~J.} \bibnamefont{Scalapino}},
  \bibnamefont{and} \bibinfo{author}{\bibfnamefont{S.~R.} \bibnamefont{White}},
  \bibinfo{journal}{Proceedings of the National Academy of Sciences}
  \textbf{\bibinfo{volume}{118}}, \bibinfo{pages}{e2109978118}
  (\bibinfo{year}{2021}),
  \eprint{https://www.pnas.org/doi/pdf/10.1073/pnas.2109978118},
  \urlprefix\url{https://www.pnas.org/doi/abs/10.1073/pnas.2109978118}.

\bibitem[{\citenamefont{Sylju\aa{}sen and Sandvik}(2002)}]{PhysRevE.66.046701}
\bibinfo{author}{\bibfnamefont{O.~F.} \bibnamefont{Sylju\aa{}sen}}
  \bibnamefont{and} \bibinfo{author}{\bibfnamefont{A.~W.}
  \bibnamefont{Sandvik}}, \bibinfo{journal}{Phys. Rev. E}
  \textbf{\bibinfo{volume}{66}}, \bibinfo{pages}{046701}
  (\bibinfo{year}{2002}),
  \urlprefix\url{https://link.aps.org/doi/10.1103/PhysRevE.66.046701}.

\bibitem[{foo({\natexlab{a}})}]{foot6}
\bibinfo{note}{{The calculation of SSE QMC is performed with IsingMonteCarlo
  package at https://github.com/Renmusxd/IsingMonteCarlo}}.

\bibitem[{\citenamefont{Ran}(2020)}]{PhysRevA.101.032310}
\bibinfo{author}{\bibfnamefont{S.-J.} \bibnamefont{Ran}},
  \bibinfo{journal}{Phys. Rev. A} \textbf{\bibinfo{volume}{101}},
  \bibinfo{pages}{032310} (\bibinfo{year}{2020}),
  \urlprefix\url{https://link.aps.org/doi/10.1103/PhysRevA.101.032310}.

\bibitem[{foo({\natexlab{b}})}]{foot5}
\bibinfo{note}{{By contracting the additional disentangler layer in
  Fig.~\ref{DMRG_arrangements} (c) into the original MPS wave-function, the
  effective bond dimension for the resulting MPS is much larger than the
  original MPS. Thus, FAMPS is a more entangled wave-function than MPS.}}

\bibitem[{\citenamefont{Evenbly and
  Vidal}(2009{\natexlab{b}})}]{PhysRevB.79.144108}
\bibinfo{author}{\bibfnamefont{G.}~\bibnamefont{Evenbly}} \bibnamefont{and}
  \bibinfo{author}{\bibfnamefont{G.}~\bibnamefont{Vidal}},
  \bibinfo{journal}{Phys. Rev. B} \textbf{\bibinfo{volume}{79}},
  \bibinfo{pages}{144108} (\bibinfo{year}{2009}{\natexlab{b}}),
  \urlprefix\url{https://link.aps.org/doi/10.1103/PhysRevB.79.144108}.

\bibitem[{\citenamefont{Xiang et~al.}(2001)\citenamefont{Xiang, Lou, and
  Su}}]{PhysRevB.64.104414}
\bibinfo{author}{\bibfnamefont{T.}~\bibnamefont{Xiang}},
  \bibinfo{author}{\bibfnamefont{J.}~\bibnamefont{Lou}}, \bibnamefont{and}
  \bibinfo{author}{\bibfnamefont{Z.}~\bibnamefont{Su}}, \bibinfo{journal}{Phys.
  Rev. B} \textbf{\bibinfo{volume}{64}}, \bibinfo{pages}{104414}
  (\bibinfo{year}{2001}),
  \urlprefix\url{https://link.aps.org/doi/10.1103/PhysRevB.64.104414}.

\bibitem[{foo({\natexlab{c}})}]{foot1}
\bibinfo{note}{We can still easily find vertical cut which crosses only one
  bond, but the number of these cuts is smaller.}

\bibitem[{\citenamefont{Singh et~al.}(2011)\citenamefont{Singh, Pfeifer, and
  Vidal}}]{PhysRevB.83.115125}
\bibinfo{author}{\bibfnamefont{S.}~\bibnamefont{Singh}},
  \bibinfo{author}{\bibfnamefont{R.~N.~C.} \bibnamefont{Pfeifer}},
  \bibnamefont{and} \bibinfo{author}{\bibfnamefont{G.}~\bibnamefont{Vidal}},
  \bibinfo{journal}{Phys. Rev. B} \textbf{\bibinfo{volume}{83}},
  \bibinfo{pages}{115125} (\bibinfo{year}{2011}),
  \urlprefix\url{https://link.aps.org/doi/10.1103/PhysRevB.83.115125}.

\bibitem[{\citenamefont{Choo et~al.}(2019)\citenamefont{Choo, Neupert, and
  Carleo}}]{PhysRevB.100.125124}
\bibinfo{author}{\bibfnamefont{K.}~\bibnamefont{Choo}},
  \bibinfo{author}{\bibfnamefont{T.}~\bibnamefont{Neupert}}, \bibnamefont{and}
  \bibinfo{author}{\bibfnamefont{G.}~\bibnamefont{Carleo}},
  \bibinfo{journal}{Phys. Rev. B} \textbf{\bibinfo{volume}{100}},
  \bibinfo{pages}{125124} (\bibinfo{year}{2019}),
  \urlprefix\url{https://link.aps.org/doi/10.1103/PhysRevB.100.125124}.

\bibitem[{\citenamefont{Nomura and Imada}(2021)}]{PhysRevX.11.031034}
\bibinfo{author}{\bibfnamefont{Y.}~\bibnamefont{Nomura}} \bibnamefont{and}
  \bibinfo{author}{\bibfnamefont{M.}~\bibnamefont{Imada}},
  \bibinfo{journal}{Phys. Rev. X} \textbf{\bibinfo{volume}{11}},
  \bibinfo{pages}{031034} (\bibinfo{year}{2021}),
  \urlprefix\url{https://link.aps.org/doi/10.1103/PhysRevX.11.031034}.

\bibitem[{\citenamefont{Hu et~al.}(2013)\citenamefont{Hu, Becca, Parola, and
  Sorella}}]{PhysRevB.88.060402}
\bibinfo{author}{\bibfnamefont{W.-J.} \bibnamefont{Hu}},
  \bibinfo{author}{\bibfnamefont{F.}~\bibnamefont{Becca}},
  \bibinfo{author}{\bibfnamefont{A.}~\bibnamefont{Parola}}, \bibnamefont{and}
  \bibinfo{author}{\bibfnamefont{S.}~\bibnamefont{Sorella}},
  \bibinfo{journal}{Phys. Rev. B} \textbf{\bibinfo{volume}{88}},
  \bibinfo{pages}{060402} (\bibinfo{year}{2013}),
  \urlprefix\url{https://link.aps.org/doi/10.1103/PhysRevB.88.060402}.

\bibitem[{\citenamefont{Paeckel et~al.}(2019)\citenamefont{Paeckel, K\"ohler,
  Swoboda, Manmana, Schollw\"ock, and Hubig}}]{PAECKEL2019167998}
\bibinfo{author}{\bibfnamefont{S.}~\bibnamefont{Paeckel}},
  \bibinfo{author}{\bibfnamefont{T.}~\bibnamefont{K\"ohler}},
  \bibinfo{author}{\bibfnamefont{A.}~\bibnamefont{Swoboda}},
  \bibinfo{author}{\bibfnamefont{S.~R.} \bibnamefont{Manmana}},
  \bibinfo{author}{\bibfnamefont{U.}~\bibnamefont{Schollw\"ock}},
  \bibnamefont{and} \bibinfo{author}{\bibfnamefont{C.}~\bibnamefont{Hubig}},
  \bibinfo{journal}{Annals of Physics} \textbf{\bibinfo{volume}{411}},
  \bibinfo{pages}{167998} (\bibinfo{year}{2019}), ISSN
  \bibinfo{issn}{0003-4916},
  \urlprefix\url{https://www.sciencedirect.com/science/article/pii/S0003491619302532}.

\bibitem[{\citenamefont{Ba\~nuls et~al.}(2009)\citenamefont{Ba\~nuls, Hastings,
  Verstraete, and Cirac}}]{PhysRevLett.102.240603}
\bibinfo{author}{\bibfnamefont{M.~C.} \bibnamefont{Ba\~nuls}},
  \bibinfo{author}{\bibfnamefont{M.~B.} \bibnamefont{Hastings}},
  \bibinfo{author}{\bibfnamefont{F.}~\bibnamefont{Verstraete}},
  \bibnamefont{and} \bibinfo{author}{\bibfnamefont{J.~I.} \bibnamefont{Cirac}},
  \bibinfo{journal}{Phys. Rev. Lett.} \textbf{\bibinfo{volume}{102}},
  \bibinfo{pages}{240603} (\bibinfo{year}{2009}),
  \urlprefix\url{https://link.aps.org/doi/10.1103/PhysRevLett.102.240603}.

\bibitem[{\citenamefont{White and Feiguin}(2004)}]{PhysRevLett.93.076401}
\bibinfo{author}{\bibfnamefont{S.~R.} \bibnamefont{White}} \bibnamefont{and}
  \bibinfo{author}{\bibfnamefont{A.~E.} \bibnamefont{Feiguin}},
  \bibinfo{journal}{Phys. Rev. Lett.} \textbf{\bibinfo{volume}{93}},
  \bibinfo{pages}{076401} (\bibinfo{year}{2004}),
  \urlprefix\url{https://link.aps.org/doi/10.1103/PhysRevLett.93.076401}.

\bibitem[{\citenamefont{Gray}(2018)}]{gray2018quimb}
\bibinfo{author}{\bibfnamefont{J.}~\bibnamefont{Gray}},
  \bibinfo{journal}{Journal of Open Source Software}
  \textbf{\bibinfo{volume}{3}}, \bibinfo{pages}{819} (\bibinfo{year}{2018}).

\bibitem[{foo({\natexlab{d}})}]{foot7}
\bibinfo{note}{{The SU(2) symmetry code is developed with TensorKit package at
  https://github.com/Jutho/TensorKit.jl}}.

\end{thebibliography}


\begin{thebibliography}{3}
\expandafter\ifx\csname natexlab\endcsname\relax\def\natexlab#1{#1}\fi
\expandafter\ifx\csname bibnamefont\endcsname\relax
  \def\bibnamefont#1{#1}\fi
\expandafter\ifx\csname bibfnamefont\endcsname\relax
  \def\bibfnamefont#1{#1}\fi
\expandafter\ifx\csname citenamefont\endcsname\relax
  \def\citenamefont#1{#1}\fi
\expandafter\ifx\csname url\endcsname\relax
  \def\url#1{\texttt{#1}}\fi
\expandafter\ifx\csname urlprefix\endcsname\relax\def\urlprefix{URL }\fi
\providecommand{\bibinfo}[2]{#2}
\providecommand{\eprint}[2][]{\url{#2}}

\bibitem[{\citenamefont{Evenbly and Vidal}(2009)}]{PhysRevB.79.144108}
\bibinfo{author}{\bibfnamefont{G.}~\bibnamefont{Evenbly}} \bibnamefont{and}
  \bibinfo{author}{\bibfnamefont{G.}~\bibnamefont{Vidal}},
  \bibinfo{journal}{Phys. Rev. B} \textbf{\bibinfo{volume}{79}},
  \bibinfo{pages}{144108} (\bibinfo{year}{2009}),
  \urlprefix\url{https://link.aps.org/doi/10.1103/PhysRevB.79.144108}.

\bibitem[{\citenamefont{Felser et~al.}(2021)\citenamefont{Felser, Notarnicola,
  and Montangero}}]{PhysRevLett.126.170603}
\bibinfo{author}{\bibfnamefont{T.}~\bibnamefont{Felser}},
  \bibinfo{author}{\bibfnamefont{S.}~\bibnamefont{Notarnicola}},
  \bibnamefont{and}
  \bibinfo{author}{\bibfnamefont{S.}~\bibnamefont{Montangero}},
  \bibinfo{journal}{Phys. Rev. Lett.} \textbf{\bibinfo{volume}{126}},
  \bibinfo{pages}{170603} (\bibinfo{year}{2021}),
  \urlprefix\url{https://link.aps.org/doi/10.1103/PhysRevLett.126.170603}.

\bibitem[{\citenamefont{Singh and Vidal}(2012)}]{PhysRevB.86.195114}
\bibinfo{author}{\bibfnamefont{S.}~\bibnamefont{Singh}} \bibnamefont{and}
  \bibinfo{author}{\bibfnamefont{G.}~\bibnamefont{Vidal}},
  \bibinfo{journal}{Phys. Rev. B} \textbf{\bibinfo{volume}{86}},
  \bibinfo{pages}{195114} (\bibinfo{year}{2012}),
  \urlprefix\url{https://link.aps.org/doi/10.1103/PhysRevB.86.195114}.

\end{thebibliography}

\end{document}


\title{Supplementary Materials for ``Augmenting Density Matrix Renormalization Group with Disentanglers''}

\author{Xiangjian Qian}
\affiliation{Key Laboratory of Artificial Structures and Quantum Control (Ministry of Education),  School of Physics and Astronomy, Shanghai Jiao Tong University, Shanghai 200240, China}

\author{Mingpu Qin} \thanks{qinmingpu@sjtu.edu.cn}
\affiliation{Key Laboratory of Artificial Structures and Quantum Control (Ministry of Education),  School of Physics and Astronomy, Shanghai Jiao Tong University, Shanghai 200240, China}

\maketitle

\begin{figure}[t]
	\includegraphics[width=80mm]{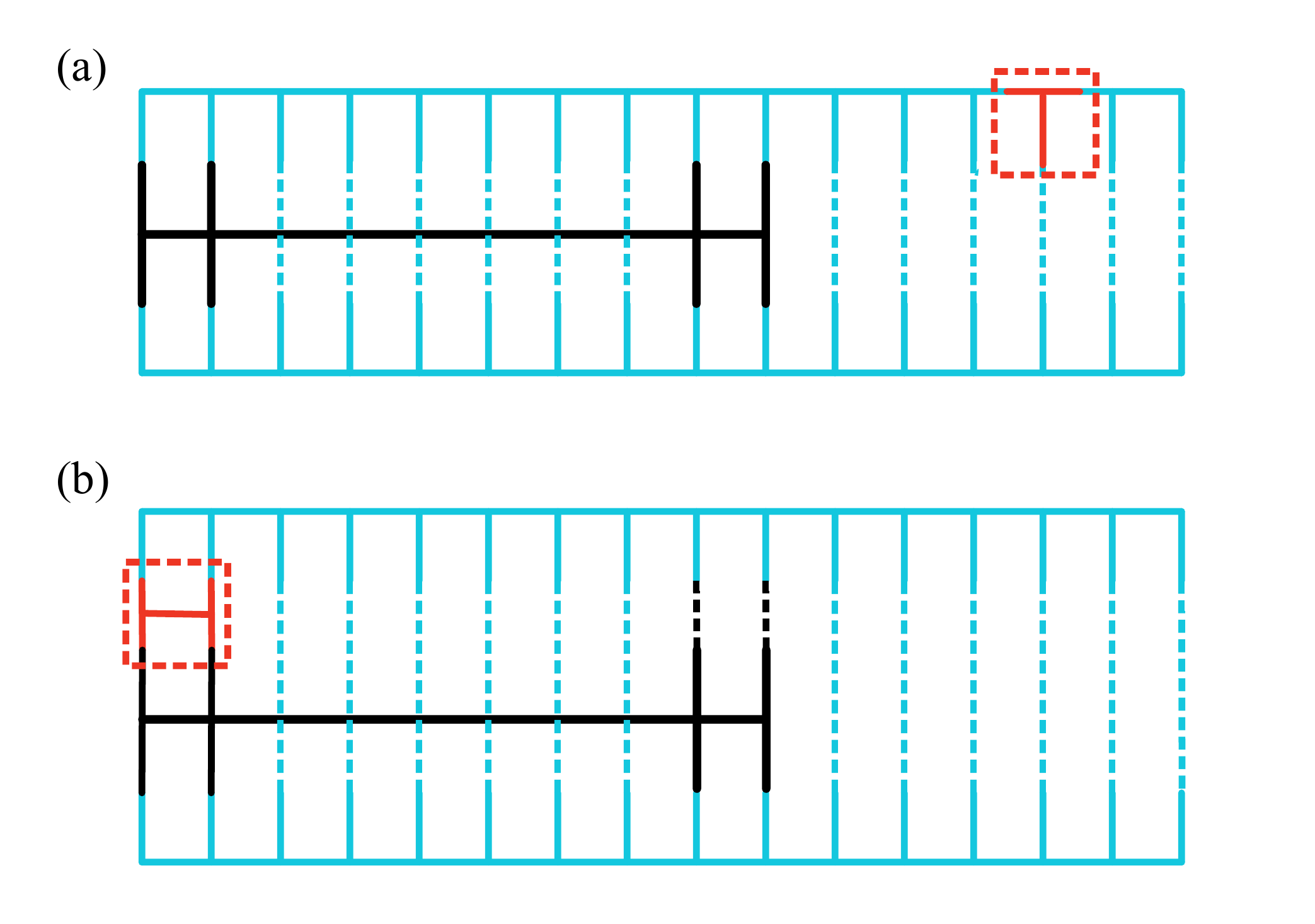}
	\caption{{(a) The environment for an MPS tensor. (b) The environment for a disentangler. The contractions of (a) and (b) can be performed with a computational cost of $O(D^3)$. The black line denotes a four-site operator. The red lines are the tensors to be optimized.}}
	\label{tensor_envir}
\end{figure}

\subsection{{Optimization of FAMPS}}
{
	We use a one dimensional representation of FAMPS (as shown in Fig.~1 (c) in the main text) to illustrate the optimization steps.
	The optimization process of FAMPS is similar as MPS which consists of the calculation of environment tensors and the solving of an eigenvalue problem. 
	The difference is that in FAMPS
	the calculation of environment is more involved because of the disentangler layer. Here, we consider a Hamiltonian which only contains two-site operators,
	\begin{equation}
		H=\sum_{ij} h_{ij}
	\end{equation}
}{We first discuss the optimization of the tensors of MPS. Because disentanglers cannot share the same site, each term of $H$ is mostly connected to two disentanglers.
	Thus, after the application of disentanglers, any two-site operator $h_{ij}$ connects with at most four sites
	\begin{equation}
		h^{\prime}_{ij}=D(u)^{\dagger} h_{ij} D(u)=u_{ik}^{\dagger} u_{jl}^{\dagger} h_{ij} u_{ik} u_{jl}=O_{ijkl}
	\end{equation}
	where $D(u)=\prod_m u_m$ represents the disentangler layer. Other disentanglers not acting on sites $i,j$ are annihilated due to $u^{\dagger}u=I$. The Hamiltonian $H$ becomes
	\begin{equation}
		H^{\prime}=D(u)^{\dagger} \sum_{ij} h_{ij} D(u)=\sum_{ij} h_{ij}^{\prime}
	\end{equation}
	where $h_{ij}^\prime$ is at most a four-site operator. So, a FAMPS with a Hamiltonian $H$ is transformed to an MPS with an effective Hamiltonian $H^\prime$. More specifically,
	\begin{equation}
		\begin{split}
			\langle \text{FAMPS}|H|\text{FAMPS}\rangle & =\langle \text{MPS}|D(u)^{\dagger}HD(u)|\text{MPS}\rangle\\
			&=\langle \text{MPS}|H^{\prime}|\text{MPS}\rangle
		\end{split}
	\end{equation}
	Thus, we can follow the procedures in MPS to optimize MPS tensors in FAMPS.
	An environment of an MPS tensor for a four-site operator is shown in Fig.~\ref{tensor_envir} (a). The contraction of it has a cost of $O(D^3)$. We need to calculate the environment for each term in the Hamiltonian $H$ to get the overall environment.}

{
	For the optimization of disentanglers,  we follow the procedure shown in Ref. \cite{PhysRevB.79.144108,PhysRevLett.126.170603} where the
	key step is also the computation of environment. Here, after the application of disentanglers, any two-site operator $h_{ij}$ is also at most a four-site operator
	\begin{equation}
		h^{\prime\prime}_{ij}=\prod_{m}  u_m^{\dagger} h_{ij} \prod_{m\neq \{ik\}} u_m=u_{ik}^{\dagger} u_{jl}^{\dagger} h_{ij} u_{jl}=O^{\prime}_{ijkl}
	\end{equation}
	where $u_{ik}$ is the disentangler we want to optimize.
	Fig.~\ref{tensor_envir} (b) shows the corresponding environment for a disentangler, which can also be contracted with a complexity of $O(D^3)$. The overall environment only involves Hamiltonian terms which are directly attached to the disentangler $u_{ik}$, which makes the optimization of disentanglers very fast. All other Hamiltonian terms contribute a constant term to the energy expectation value and are independent of $u_{ik}$ \cite{PhysRevB.79.144108}. More specifically,
	\begin{equation}
		\begin{split}
			\langle \text{FAMPS}|H|\text{FAMPS}\rangle & =\langle \text{MPS}|D(u)^{\dagger}HD(u)|\text{MPS}\rangle\\
			&=\langle \text{MPS}|\sum_{p} h_{p}^{\prime\prime} u_{ik}|\text{MPS}\rangle+C_{ik}
		\end{split}
	\end{equation}
	where $p$ represents the Hamiltonian terms that act on $u_{ik}$ and $C_{ik}=\langle \text{MPS}|\sum_{k} h_{k}^{\prime}|\text{MPS}\rangle$ with $k$ the other Hamiltonian terms not acting on $u_{ik}$.}

{
	Overall, the optimization process of FAMPS is divided into two steps: the optimization of disentanglers and the optimization of the MPS tensors. 
	For the optimization of disentanglers, the procedure is described as follows.
	\begin{itemize}
		\item[(i)]  select a disentangler $u$, calculate the environments for this disentangler from the Hamiltonian terms that are directly attached to it, add them up to get the overall environment $E$.
		\item[(ii)] perform a Singular Value Decomposition (SVD) for environment $E=USV^\dagger$, then calculate the optimized disentangler $u_{\text{new}}=-VU^\dagger$.  
		\item[(iii)] move to another disentangler and repeat (i) and (ii) until all the disentangles are optimized once. 
	\end{itemize}
	After optimizing the disentanglers for $3\sim 5$ sweeps, we move to the optimization of MPS tensors.
	The optimization of MPS tensors follows the below steps.
	\begin{itemize}
		\item[(i)]  choose an MPS tensor $T$, set the tensor $T$ as the center tensor as in MPS optimization algorithm.
		\item[(ii)] calculate the environments for this tensor $T$ from all the Hamiltonian terms and add them up to get the overall environment.
		\item[(iii)] use the Lanczos or similar algorithm to solve the eigenvalue problem for tensor T. In this step, step (ii) needs to be repeated for several times.  
		\item[(iv)] move to a next MPS tensor and repeat step (i), (ii) and (iii) until all the MPS tensors are optimized. 
	\end{itemize}
	Then, we repeat the process by optimizing disentanglers and MPS tensors alternatively until the energy is converged.}

\begin{figure}[t]
	\includegraphics[width=42mm]{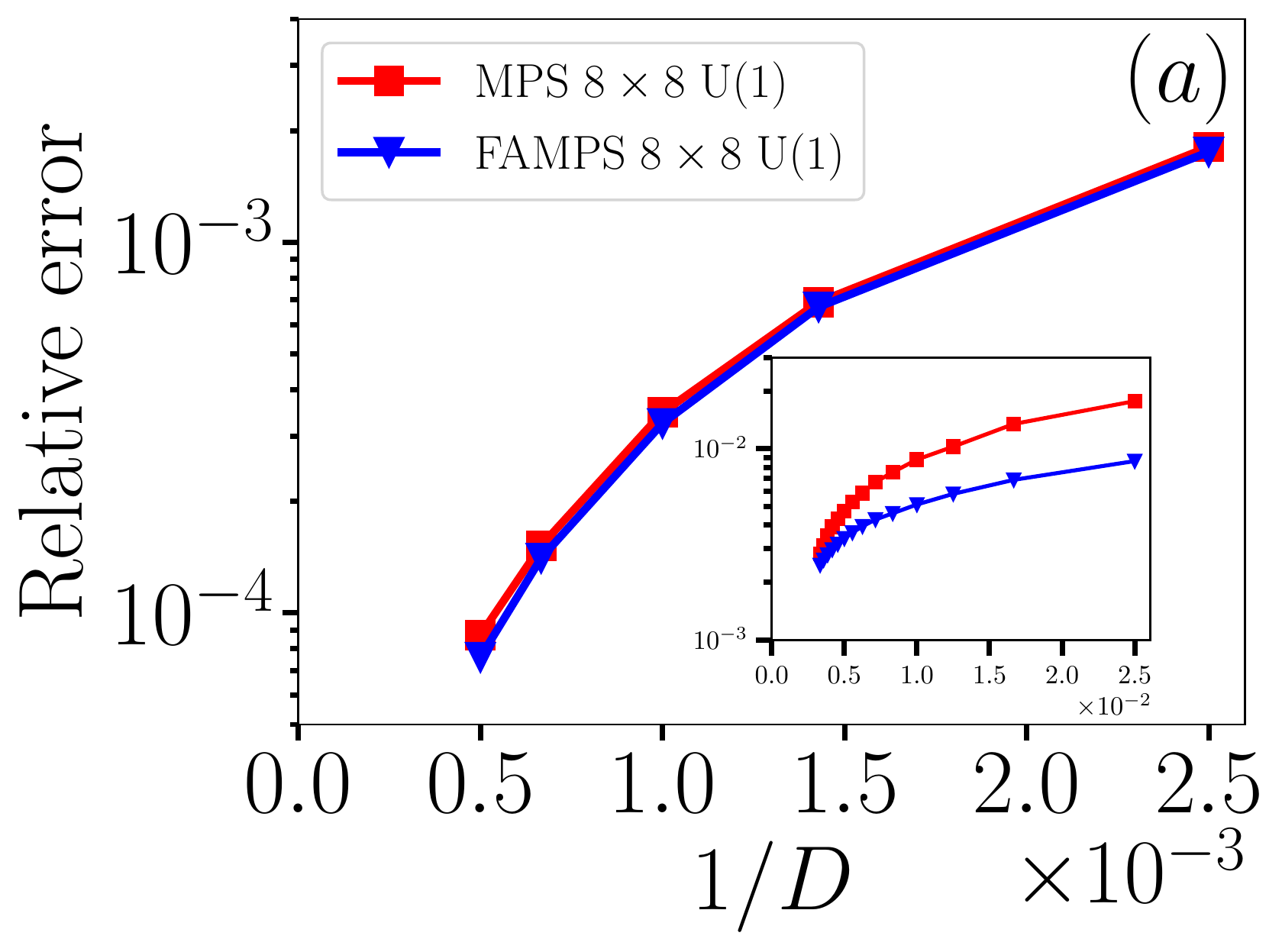}
	\includegraphics[width=42mm]{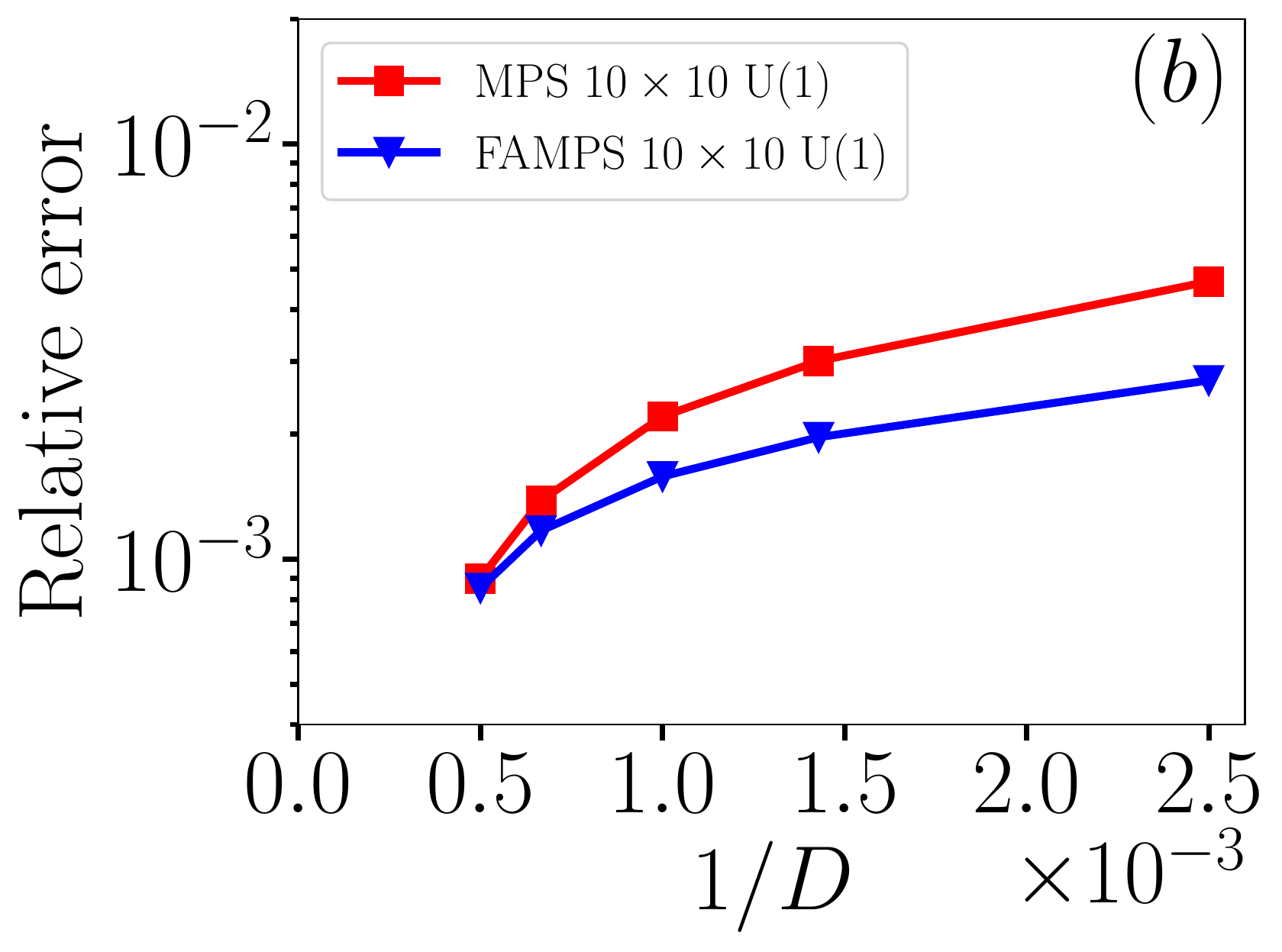}
	\includegraphics[width=42mm]{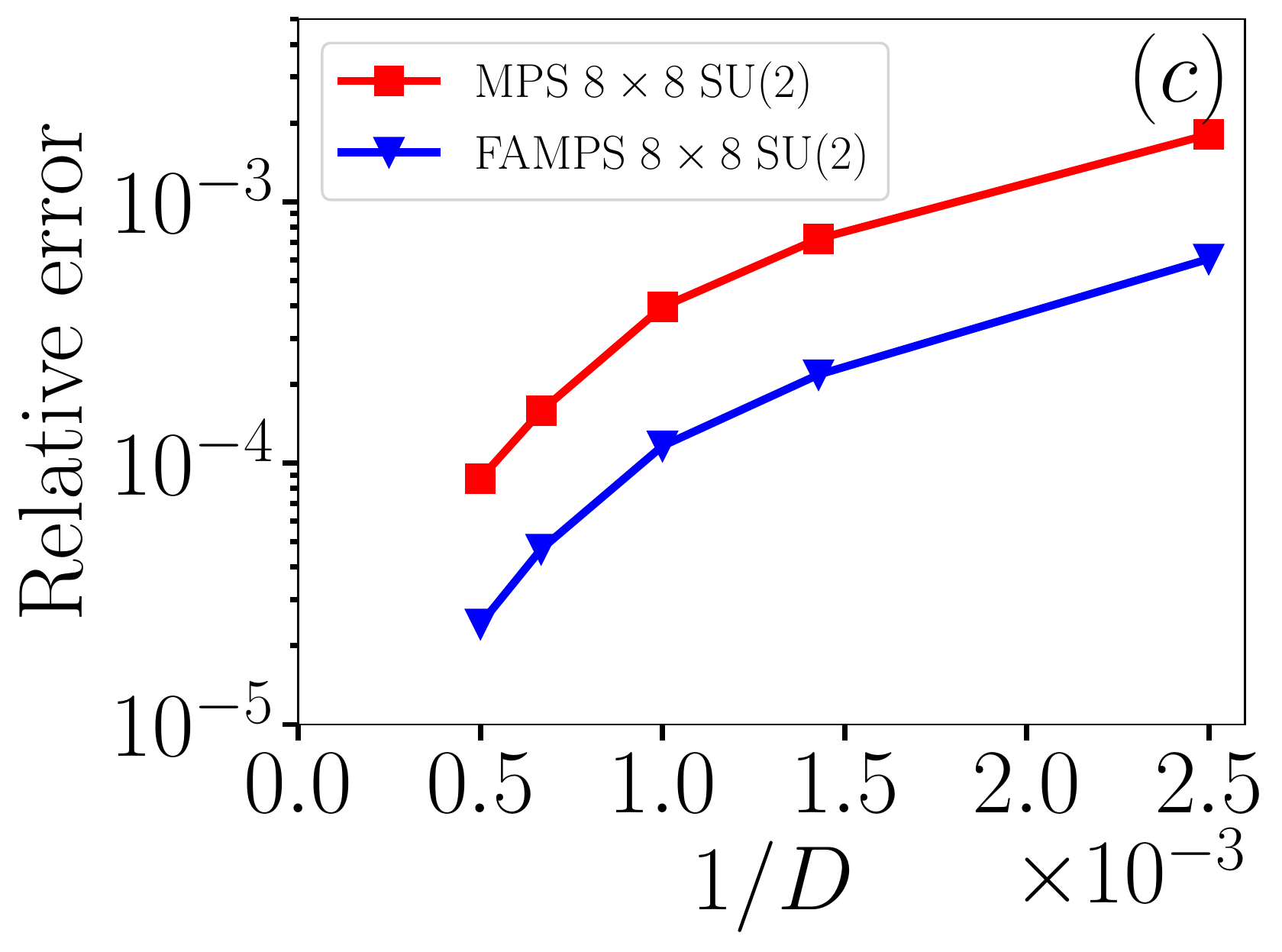}
	\includegraphics[width=42mm]{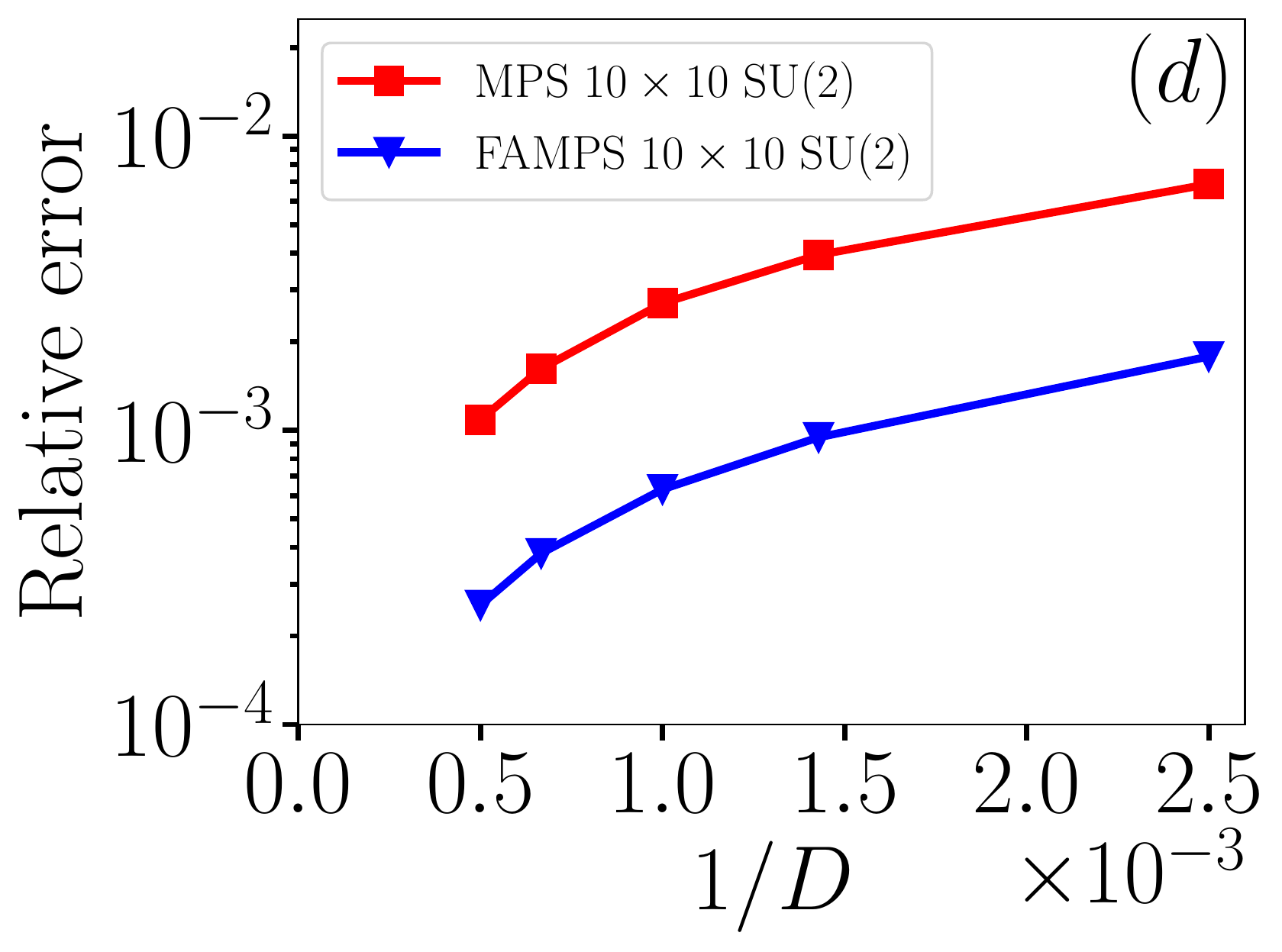}
	
	\caption{{Relative error of the ground state energy for FAMPS and MPS of the Heisenberg model for lattice sizes $8\times 8$ and $10\times 10$ with open boundary conditions. Simulations in (a) and (b) utilize U(1) symmetry and have a physical bond dimension $d=2$, while in (c) and (d), we use SU(2) symmetry and have a physical bond dimension $d=4$ by blocking two sites into one site. We can find that by allowing more free parameters in the disentanglers (in (c) and (d)), the critical bond dimension problem (in (a) and (b)) is solved. }}
	\label{Heisenberg_OBCs}
\end{figure} 

\begin{figure}[t]
	\includegraphics[width=80mm]{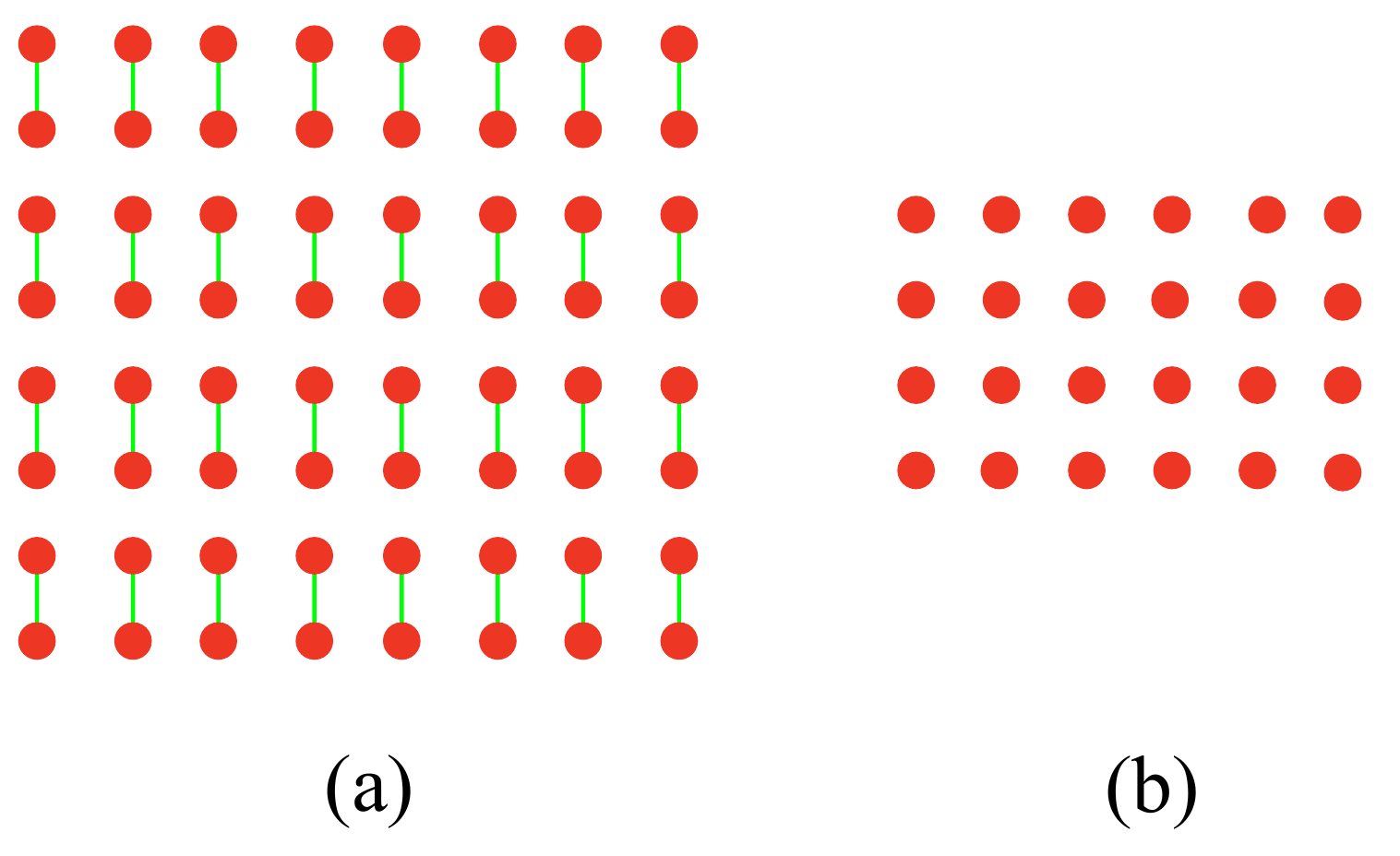}
	\caption{{To increase the number of free parameters in disentanglers for the Heisenberg model, we block two sites linked by green lines in (a) into one site.}}
	\label{blocking_sites}
\end{figure}

\subsection{{Critical Bond Dimension in FAMPS}}

{
	As mentioned in the main text, in the study of Heisenberg model with open boundary conditions we encounter the problem of critical bond dimension in FAMPS. 
	The results are shown in Fig.~\ref{Heisenberg_OBCs}. In (a) and (b), we can find that with the increase of bond dimension, there exist a critical bond dimension after which the results of FAMPS and MPS are identical. This phenomenon is related to the absence of free parameters of disentanglers. For spin $1/2$ degree of freedom,
	each site has physical dimension $d=2$ and can be labeled by quantum number $\{0,1\}$. Under U(1) symmetry, disentangler $u_{ij}^{kl}$ can be written as (we only consider real entries for $u_{ij}^{kl}$)
	\begin{equation}
		u_{ij}^{kl}=u_0 \oplus u_1 \oplus u_2
	\end{equation}
	where $u_1$ is a $2\times 2$ matrix and $u_0,u_2$ are $1\times 1$ matrices or numbers. 
	Due to the unitary property of disentangler, $u_0^2=1, u_2^2=1, u_1^{\dagger}u_1=I_{2\times 2}$.
	Thus, disentangler $u_{ij}^{kl}$ can be written as
	\begin{equation}
		\begin{split}
			u_{ij}^{kl}=&\pm 1\left(                 
			\begin{array}{cccc}   
				1 & 0 & 0 & 0\\  
				0 &  \sin \theta &  \cos \theta & 0\\
				0 & -\cos \theta & \sin \theta & 0\\
				0 & 0 & 0 &  1\\   
			\end{array}
			\right) 
		\end{split}
	\end{equation}
	Thus, disentangler $u_{ij}^{kl}$ has only one free parameter under U(1) symmetry. 
	But under SU(2) symmetry, disentangler $u_{ij}^{kl}$ is written as (using the notation in Ref. \cite{PhysRevB.86.195114})
	\begin{equation}
		u_{ij}^{kl} = (P_0 \otimes Q_{\frac{1}{2},\frac{1}{2}\rightarrow \frac{1}{2},\frac{1}{2}}^0) + \\(P_1 \otimes Q_{\frac{1}{2},\frac{1}{2}\rightarrow \frac{1}{2},\frac{1}{2}}^1) 
	\end{equation}
	where $P_0,P_1$ are numbers, $Q_{\frac{1}{2},\frac{1}{2}\rightarrow \frac{1}{2},\frac{1}{2}}^0$ and $Q_{\frac{1}{2},\frac{1}{2}\rightarrow \frac{1}{2},\frac{1}{2}}^1$ are matrix determined by Clesch-Gordan coefficients, which are $  \left(                 
	\begin{array}{cccc}   
	0 & 0 & 0 & 0\\  
	0 &  1/2 &  -1/2 & 0\\
	0 & -1/2 & 1/2 & 0\\
	0 & 0 & 0 &  0\\   
	\end{array}
	\right)$ and $
	\left(                 
	\begin{array}{cccc}   
	1  & 0 & 0 & 0\\  
	0 &  1/2 &  1/2 & 0\\
	0 & 1/2 & 1/2 & 0\\
	0 & 0 & 0 &  1\\   
	\end{array}
	\right)$ respectively. Under the restriction of $u^\dagger u=I$, $P_0^2=P_1^2=1$, which leads to $P_0=\pm 1,P_1=\pm 1$. Thus, disentangler $u_{ij}^{kl}$ has no free parameter under SU(2) symmetry. When considering Z(2) symmetry (in the transverse Ising model), disentangler $u_{ij}^{kl}$ can be written as \begin{equation}
		u_{ij}^{kl}=u_0 \oplus u_1
	\end{equation} where $u_0$, $u_1$ are $2\times 2$ matrix satisfying $u_1^{\dagger}u_1=u_2^{\dagger}u_2=I_{2\times 2}$. Then, disentangler $u_{ij}^{kl}$ can
	be represented as
	\begin{equation}
		u_{ij}^{kl}=\left(                 
		\begin{array}{cccc}   
			\sin \alpha & 0 & 0 & \cos \alpha\\  
			0 &  \sin \beta &  \cos \beta & 0\\
			0 & -\cos \beta & \sin \beta & 0\\
			-\cos \alpha & 0 & 0 &  \sin \alpha\\   
		\end{array}
		\right)
	\end{equation} Here, disentangler $u_{ij}^{kl}$ has two free parameters. Thus, in Transverse Ising model, there is no critical bond dimension as shown in Fig.~2 in
the main text.}

{
	For the U(1) symmetry imposed DMRG calculations in Fig.~\ref{Heisenberg_OBCs} (a) and (b), SU(2) symmetry is nearly restored after the critical bond dimension,
	which means the disentanglers have no 
	free parameter and FAMPS results are identical to MPS ones. To solve this problem, we can block two sites into one to enlarge the physical degrees of freedom as shown in Fig.~\ref{blocking_sites}. In this way, we increase the number of free parameters in the disentanglers. Fig.~\ref{Heisenberg_OBCs} (c) and (d) show calculations with the same models in Fig.~\ref{Heisenberg_OBCs} (a) and (b). From Fig.~\ref{Heisenberg_OBCs}, we can find that the FAMPS results are always better than MPS results and there is no critical bond dimension.}

\begin{figure}[t]
	\includegraphics[width=80mm]{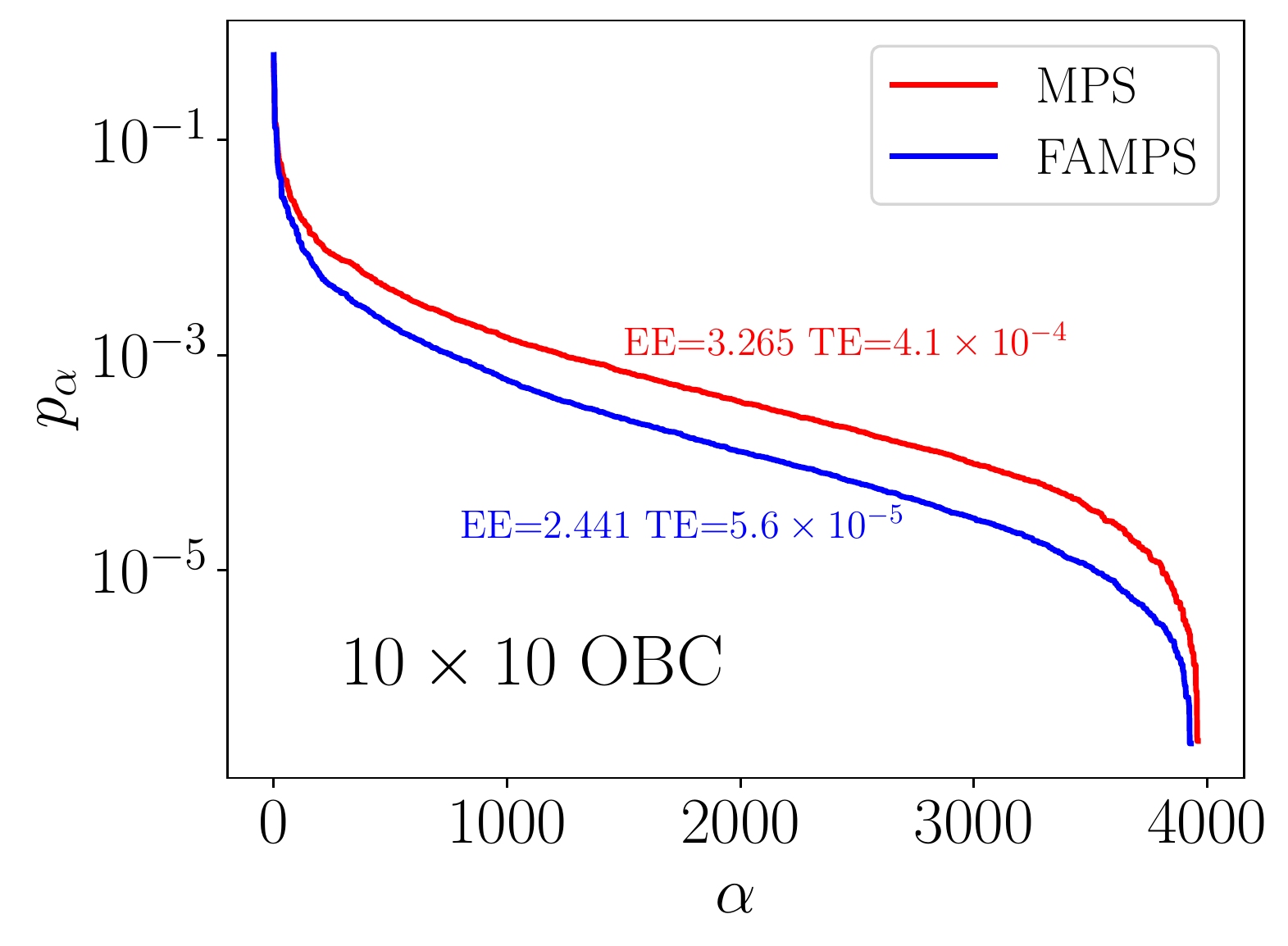}
	\caption{The distribution of the entanglement spectrum for the Heisenberg model on a $10 \times 10 $ lattice under open boundary conditions. The bond dimension for both MPS and FAMPS are $D = 1000$. The result
		for the FAMPS is for the underlying MPS (without disentanglers). We can see that
		both the entanglement entropy (EE) and truncation error (TE) are reduced in FAMPS.
	}
	\label{spectrum}
\end{figure}

\begin{figure}[t]
	\includegraphics[width=80mm]{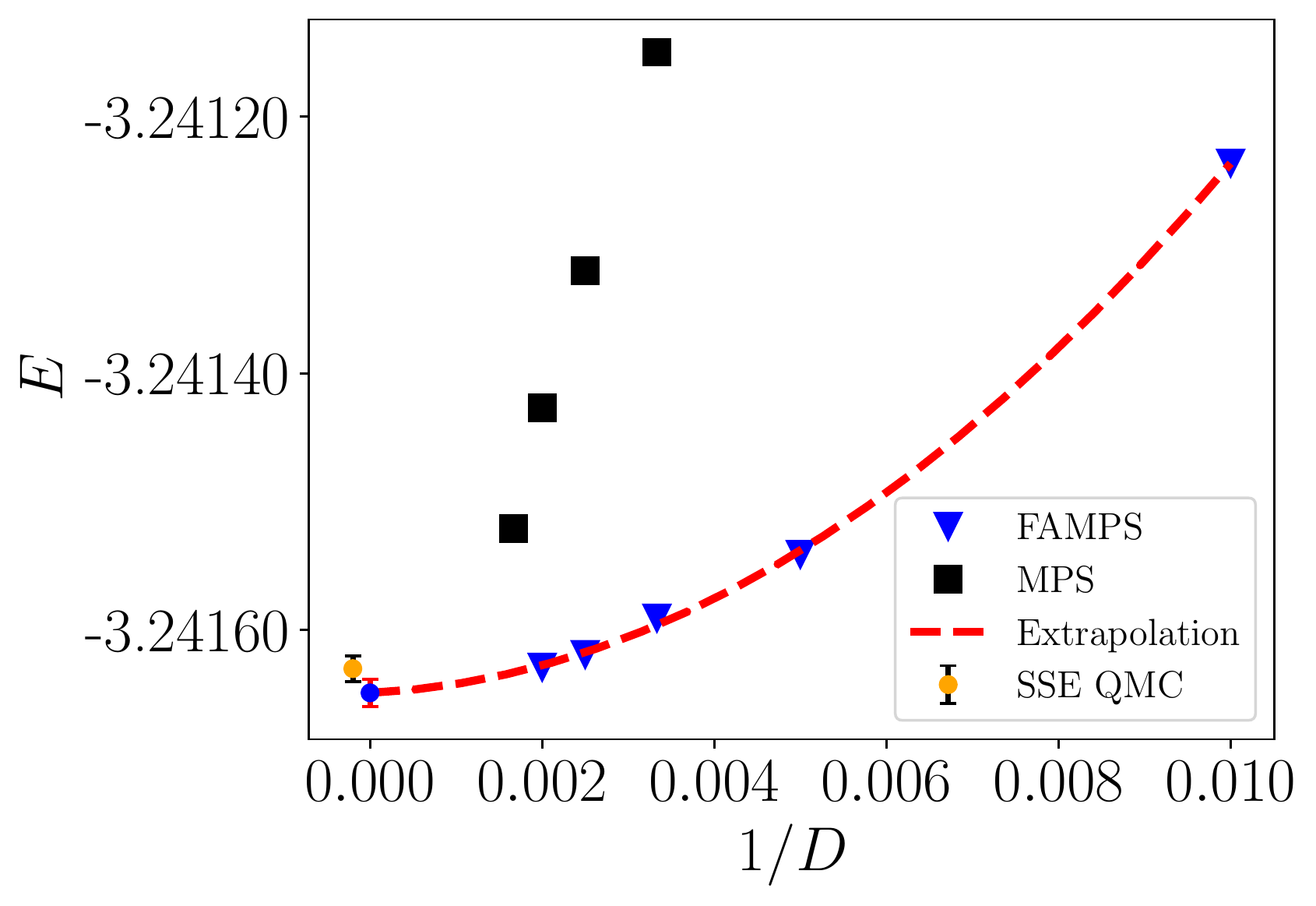}
	\caption{A quadratic extrapolation of the FAMPS energy per site of a $8 \times 8$ transverse Ising model with $\lambda = 3.05$ under periodic boundary
		conditions.
		The fit gives an extrapolated value of ground state energy per site $E=-3.24165(1)$ which agrees with the SSE QMC result $-3.24163(1)$ within the error bar.}
	\label{extra_ising}
\end{figure}

\subsection{Comparison of the entanglement spectrum}
{
In Fig.~\ref{spectrum}, we show the the distribution of the entanglement spectrum for the Heisenberg model on a $10 \times 10 $ lattice under open boundary conditions. The bond
dimension for both MPS and FAMPS are $D = 1000$. The result for the FAMPS is for the underlying MPS (without disentanglers). We can see that both the entanglement entropy (EE) and truncation error (TE) are reduced in FAMPS, which
demonstrates the effect of disentanglers to reduce the entanglement in the ground state wave-function and hence to increase the accuracy of MPS.
}

\subsection{More Results}
In Fig.~\ref{extra_ising}, we show the extrapolation of FAMPS energy with bond dimension for the transverse Ising model with periodic boundary conditions and $\lambda = 3.05$ on a $8\times 8 $ lattice. {A quadratic fit gives the extrapolated value of ground state energy per site $E=-3.24165(1)$ which agrees with the SSE QMC result $-3.24163(1)$ within the error bar}.

\bibliography{FAMPS_SM}